\documentclass[aps,prb,reprint,superscriptaddress,nofootinbib]{revtex4-1}

\usepackage{mathtools}
\usepackage{amssymb,amsmath}
\usepackage{graphicx}
\usepackage{xspace}
\usepackage{tabularx}
\usepackage{algorithm}
\usepackage{algpseudocode}

\newcommand{\vectorspace}[2]{\ensuremath{\mathbb{#1}^{#2}}\xspace}
\newcommand{\argmin}{\operatornamewithlimits{argmin}}
\newcommand{\set}[1]{\ensuremath{\mathbf{#1}}}
\newcommand{\norm}[1]{\ensuremath{\left|\left|{#1}\right|\right|}}
\newcommand{\nuclearnorm}[1]{\ensuremath{\left|\left|{#1}\right|\right|_*}}
\newcommand{\frobnorm}[1]{\ensuremath{\left|\left|{#1}\right|\right|_F}}
\newcommand{\frobproduct}[1]{\ensuremath{\left\langle {#1} \right\rangle_F}}
\DeclareMathOperator{\Tr}{Tr}
\DeclareMathOperator{\diag}{diag}
\newcommand{\slthree}{\ensuremath{SL_3 \left( \vectorspace{Z}{} \right)}\xspace}

\newcommand{\deformation}{\set{F}}

\newcommand{\stretchtensor}{\set{U}}
\newcommand{\unimodular}{\set{L}}

\newcommand{\Bstrut}[1]{\rule[- #1 ex]{0pt}{0pt}}   
\newcommand{\figtext}[1]{\footnotesize\textbf{#1}}

\newcounter{varequation}
\makeatletter
\newcommand{\vareqnum}{\refstepcounter{varequation}\textup{\tagform@{P\thevarequation}}}
\makeatother
\newcommand{\vareqref}[1]{(P\ref{#1})}

\begin{document}


\title{Minimum-Strain Symmetrization of Bravais Lattices}



\author{Peter M.~Larsen}
\email{pmla@mit.edu}
\author{Edward L.~Pang}
\affiliation{Department of Materials Science and Engineering, Massachusetts Institute of Technology, Cambridge, MA 02139, USA}
\author{Pablo A. Parrilo}
\affiliation{Department of Electrical Engineering and Computer Science, Massachusetts Institute of Technology, Cambridge, MA 02139, USA}
\author{Karsten W.~Jacobsen}
\affiliation{Center for Atomic-scale Materials Design (CAMD), Department of Physics, Technical University of Denmark, 2800 Kongens Lyngby, Denmark}

\date{\today}

\begin{abstract}
Bravais lattices are the most fundamental building blocks of crystallography.
They are classified into groups according to their translational, rotational, and inversion symmetries.
In computational analysis of Bravais lattices, fulfilment of symmetry conditions is usually determined by analysis of the metric tensor, using either a numerical tolerance to produce a binary (i.e.~yes or no) classification, or a distance function which quantifies the deviation from an ideal lattice type.  The metric tensor, though, is not scale-invariant, which complicates the choice of threshold and the interpretation of the distance function.
Here, we quantify the distance of a lattice from a target Bravais class using strain.  For an arbitrary lattice, we find the minimum-strain transformation needed to fulfil the symmetry conditions of a desired Bravais lattice type; 
the norm of the strain tensor is used to quantify the degree of symmetry breaking.
The resulting distance is invariant to scale and rotation, and is a physically intuitive quantity.
By symmetrizing to all Bravais classes, each lattice can be placed in a 14 dimensional space, which we use to create a map of the space of Bravais lattices and the transformation paths between them.  A software implementation is available online under a permissive license.
\end{abstract}


\maketitle

\section{Introduction}

Many properties of a crystalline material are governed by the geometry of its Bravais lattice, and changes in the Bravais lattice, both static and dynamic.
For example, modification of the Bravais geometry by the imposition of hydrostatic and/or shear strains forms the basis of \emph{elastic strain engineering}~\cite{lu2014ese}, a field whose successes include improved photoluminescence and electronic spectra in semiconductors~\cite{suess2013germanium, minamisawa2012nanowires, shi2019elasticstrain},
vibrational properties in micromechanical oscillators~\cite{ghadimi764oscillator}, and magnetic properties in multiferroics~\cite{fennie2006ferroic}.  
Alternatively, the design of materials which can sustain repeated cyclic changes in the Bravais lattice geometry is the fundamental aim of shape memory alloy research~\cite{otsuka1999shape, auricchio1997sma, auricchio1997smamacro}.

In addition to the practical consequences for materials engineering, there is a fundamental theoretical interest in the classification of Bravais lattices.  This has motivated the development of many algorithms for lattice analysis.  A traditional lattice analysis method is straightforward to describe:
\begin{enumerate}
\item Find a canonical description of the lattice.
\item Classify the lattice according to the symmetries of its cell parameters.  
\end{enumerate}
The most commonly used canonical description is the \emph{Niggli-reduced}~\cite{niggli1928krystallographische, dewolff2006reducedbases} form of the lattice, which is defined using the metric tensor.  For a unit cell with lattice vectors $\vec{a}$, $\vec{b}$, and $\vec{c}$, the metric tensor is given by
\begin{equation}
\set{G}
=
\begin{bmatrix}
g_{11} & g_{12} & g_{13}\\
g_{12} & g_{22} & g_{23}\\
g_{13} & g_{23} & g_{33}\\
\end{bmatrix}
=
\begin{bmatrix}
  \vec{a} \cdot \vec{a}
& \vec{a} \cdot \vec{b}
& \vec{a} \cdot \vec{c}\\
  \vec{a} \cdot \vec{b}
& \vec{b} \cdot \vec{b}
& \vec{b} \cdot \vec{c}\\
  \vec{a} \cdot \vec{c}
& \vec{b} \cdot \vec{c}
& \vec{c} \cdot \vec{c}
\end{bmatrix}
\label{eq:metric_tensor}
\end{equation}
Niggli reduction specifies a set of constraints which the metric tensor must satisfy.  After reduction, the Bravais class of a lattice can be determined by inspecting the elements of its metric tensor.  For example, the Niggli-reduced form of the primitive cubic lattice is the same as the conventional setting, where the metric tensor satisfies
\begin{subequations}
\begin{align}
g_{11} = g_{22} = g_{33}\\
g_{23} = g_{13} = g_{12} = 0
\end{align}
\label{eq:cubic_exact}
\end{subequations}
Similar conditions can be specified for each of the Bravais lattice types.  Whilst constraints of this form provide a clean theoretical description of Bravais classification, practical classification is complicated by the presence of noise.  Lattice parameters which have been determined by experiment are subject to measurement errors, and even lattice parameters obtained from computer simulations are subject to numerical errors resulting from the use of floating-point arithmetic.  Both of these sources of error mean that symmetry conditions are rarely fulfilled exactly.

Noise is treated in one of two ways, using either a tolerance parameter or a distance calculation.  A tolerance parameter, $\epsilon$, is used to permit approximate, rather than exact, fulfilment of symmetry conditions.  In this case, the
cubic symmetry conditions in Equation~\eqref{eq:cubic_exact} are relaxed:
\begin{subequations}
\begin{align}
\left| g_{11} - g_{22} \right| &\leq \epsilon&
\left| g_{11} - g_{33} \right| &\leq \epsilon&
\left| g_{22} - g_{33} \right| &\leq \epsilon\\
\left| g_{23} \right| &\leq \epsilon&
\left| g_{13} \right| &\leq \epsilon&
\left| g_{12} \right| &\leq \epsilon
\end{align}
\label{eq:cubic_epsilon}
\end{subequations}
The tolerance parameter effectively defines a boundary between symmetry classes.  
Exactly where this boundary should lie is decided by the user, but involves a degree of arbitrariness: given a sufficiently large number of structures, there will be lattices on either side of any boundary, whose classification is dependent on the exact choice of threshold.  This use of a tolerance parameter is the approach used by Grosse-Kunstleve~\cite{grossekunstleve2004stableniggli} and in the lattice determination component of the \texttt{Spglib}~\cite{togo2018spglib} space group classification library.

To overcome the need for an \emph{a priori} selection of a tolerance parameter, a distance can be calculated, which effectively measures the degree of constraint violation:
\begin{subequations}
\begin{align}
d\left(G_M\right) = 
&\left| g_{11} - g_{22} \right| +
\left| g_{11} - g_{33} \right| +
\left| g_{22} - g_{33} \right|\\
&+ \left| g_{23} \right|
+ \left| g_{13} \right|
+ \left| g_{12} \right|
\end{align}
\label{eq:cubic_distance}
\end{subequations}
By measuring the distances from each Bravais class, the choice of distance threshold can be made \emph{a posteriori}, and informed by the available options.  A variety of distance functions on the metric tensor have been used in the literature~\cite{andrews1988metricdistance, macicek1992blaf, oishitomiyasu2012distance, andrews2015bgaol}.  

The commonality of existing methods is the use of the metric tensor.  Whilst the metric tensor is a rotationally invariant description of a lattice basis, it is not invariant to scale, which complicates the choice of threshold and the interpretation of the distance function.

In this work we determine the distance of a lattice from a chosen Bravais type using strain.
The problem we aim to solve is the following: given an observed lattice which we which to symmetrize, find the closest lattice in the target Bravais class.  The distance is determined by the minimum strain needed to elastically deform a lattice such that the symmetry conditions of the target Bravais class are satisfied.
Strain has the advantage of being a rotationally invariant and physically intuitive quantity, and can be easily made scale-invariant.  Furthermore, since the transformation between any two lattices is a linear map, strain also represents the most natural distance measure on lattices.
In addition to lattice classification, the other immediate application of this work is simply to symmetrize a lattice in well-defined manner.

The rest of this article is structured as follows:
we define a similarity measure for fixed Bravais lattices using the strain tensor in section~\ref{sec:similarity}.
We extend this to variable lattices in section~\ref{sec:symmetry_breaking}, which enables us to find minimum-strain symmetrizations.
The symmetrization method is illustrated in section~\ref{sec:applications}, and we use it to present a map of the Bravais lattices in section~\ref{sec:cartography}.
Concluding remarks are given in section~\ref{sec:conclusion}.

\section{Quantification of Lattice Similarity}
\label{sec:similarity}

The symmetrization procedure we develop in this work is based on quantification of the deformation from one lattice to another.  In this section we describe the necessary deformation theory, which is a cornerstone of continuum mechanics.

\subsection{Lattice Basis Comparison}

A Bravais lattice, $\Lambda_A$, is described by a lattice basis, $\set{A} \in \vectorspace{R}{3 \times 3}$, which consists of three lattice vectors, described by the columns of $\set{A}$.  The lattice points lie at all integer combinations of the lattice basis vectors, that is, a lattice point $\vec{p}$ satisfies
$\vec{p} = \set{A}\vec{h}$ where $\vec{h} \in \vectorspace{Z}{3}$.
Here we wish to compare two lattices geometrically.

Let $\Lambda_A$ and $\Lambda_B$ be Bravais lattices, with lattice bases $\set{A} \in \vectorspace{R}{3 \times 3}$ and $\set{B} \in \vectorspace{R}{3 \times 3}$. 
We will compare the bases by quantifying the minimum deformation necessary to map one basis onto the other.
%
There is a linear map, $\deformation \in \vectorspace{R}{3 \times 3}$, also called the \emph{deformation gradient}, which exactly transforms \set{B} into \set{A}:
\begin{equation}
\set{A} = \deformation\set{B}
\label{eq:linear_map}
\end{equation}
We can remove the rotational dependence of $\set{F}$ with the use of a polar decomposition:
\begin{equation}
\stretchtensor = \sqrt{\deformation^T \deformation}
\end{equation}
Here $\stretchtensor$ is a symmetric matrix containing pure stretches only, known as the \emph{right stretch tensor}.  The Doyle-Ericksen strain tensors\cite{doyle1956nonlinearelasticity} (or sometimes, the Seth-Hill strain tensors~\cite{seth1961generalized,hill1979tensor}) are a generalized class of strain tensors of the form
\begin{equation}
\set{E} = \frac{1}{m} \left( \stretchtensor^m - \set{I} \right)
=
\begin{bmatrix}
\epsilon_{11} & \epsilon_{12} & \epsilon_{13}\\
\epsilon_{12} & \epsilon_{22} & \epsilon_{23}\\
\epsilon_{13} & \epsilon_{23} & \epsilon_{33}\\
\end{bmatrix}
\label{eq:doyle}
\end{equation}
where $m$ typically takes an integer value.  In this work we choose $m=1$, in which case $\set{E}$ is the \emph{Biot} strain tensor.

Combining Equations~(\ref{eq:linear_map})-(\ref{eq:doyle}), we define a distance function which quantifies the strain tensor norm:
\begin{equation}
d \left(\set{A}, \set{B} \right)
= \frobnorm{ \set{E} }
= \frobnorm{
\sqrt{ \set{B}^{-T}\set{A}^T \set{A} \set{B}^{-1} }
- \set{I} }
\label{eq:distance_strain}
\end{equation}
where $\frobnorm{\set{X}} = \sqrt{\sum\limits_{i,j} X_{ij}^2}$ is the Frobenius norm.

The distance function is invariant to orthogonal transformations of \set{A} and \set{B} (trivially for $\set{A}$, and by the spectral theorem for $\set{B}$).
To provide some intuition for the distance function, we can express it in terms of the principal stretches, $\nu_i$, of $\deformation$:
\begin{equation}
d \left(\set{A}, \set{B} \right) = \sqrt{ 
\left( \nu_1 - 1 \right)^2 + \left( \nu_2 - 1 \right)^2 + \left( \nu_3 - 1 \right)^2
}
\end{equation}
The principal stretches are equal to the eigenvalues of $\stretchtensor$, and the distance is equal to the norm of the principal strains.
Alternatively, we can use the relationship between the polar decomposition and the singular value decomposition (SVD) to express the distance function in terms of the singular values, $\sigma_i$, of $\deformation$:
\begin{equation}
d \left(\set{A}, \set{B} \right) = \sqrt{ 
\left( \sigma_1 - 1 \right)^2
+ \left( \sigma_2 - 1 \right)^2
+ \left( \sigma_3 - 1 \right)^2 }
\label{eq:similarity_fixed_svd}
\end{equation}
The principal stretches and singular values are in fact identical and differ only in their origin and interpretation; the former formulation is from continuum mechanics and has a physical meaning, the latter is a more general formulation from linear algebra which we use in this work.

The distance function shown here is described in further detail by Koumatos and Muehlemann~\cite{koumatos2015strain}, who also discuss different choices of $m$.

{
\setlength{\fboxsep}{0pt}
\setlength{\fboxrule}{1pt}

\begin{figure}
\centering
\begin{tabular}{ccc}
\fbox{\includegraphics[width=0.31\columnwidth]{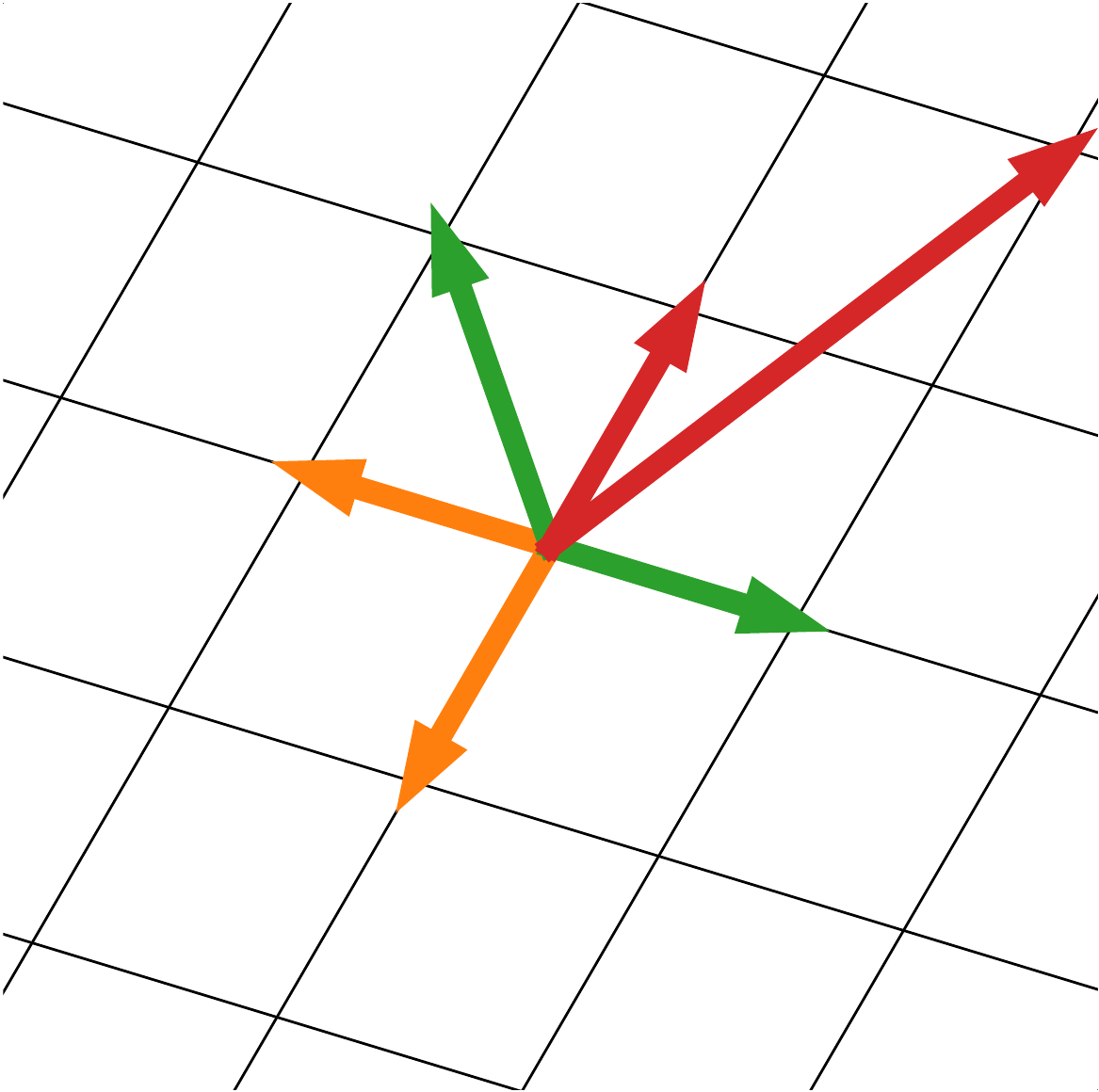}}
& \fbox{\includegraphics[width=0.31\columnwidth]{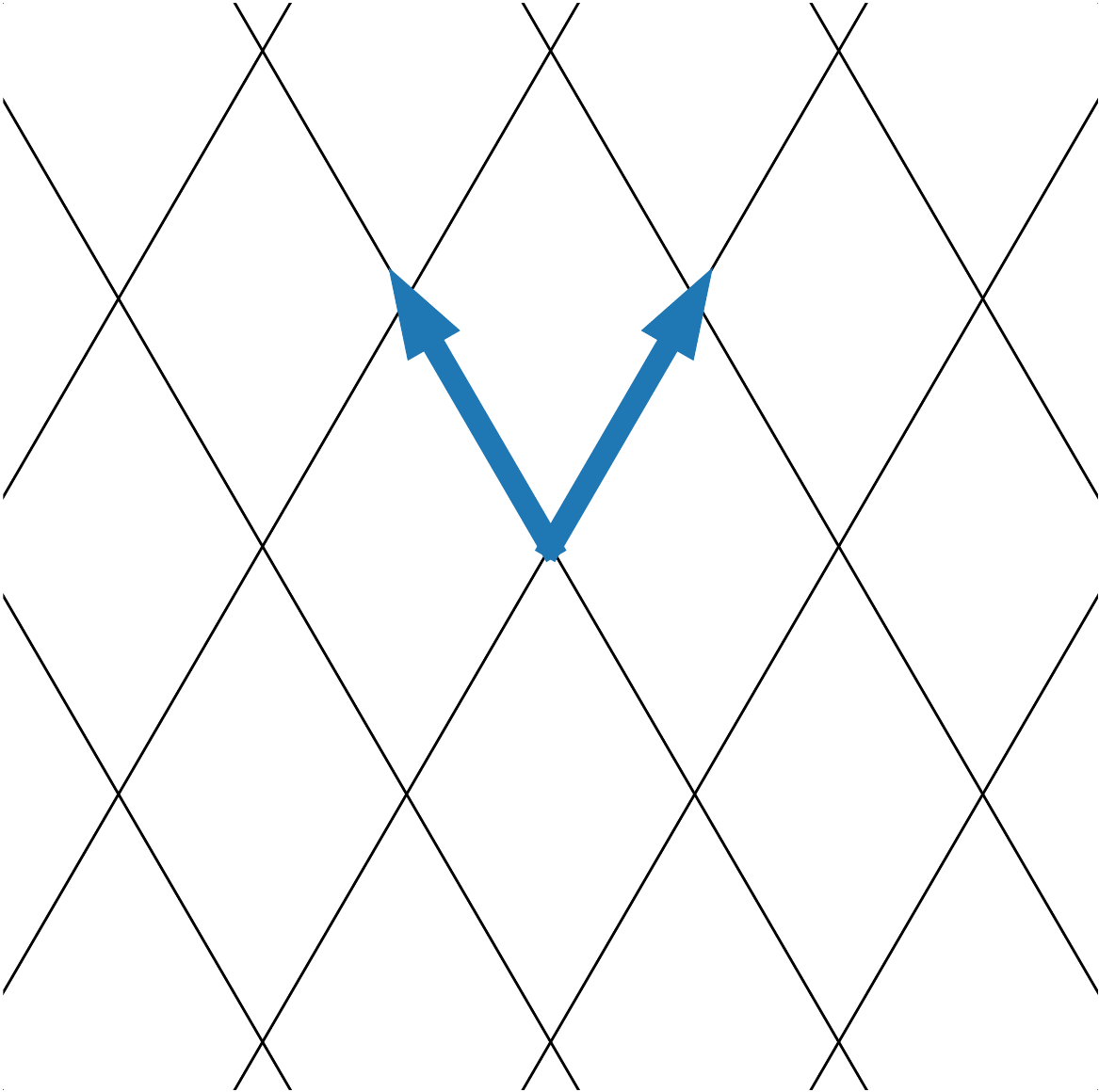}}
& \fbox{\includegraphics[width=0.31\columnwidth]{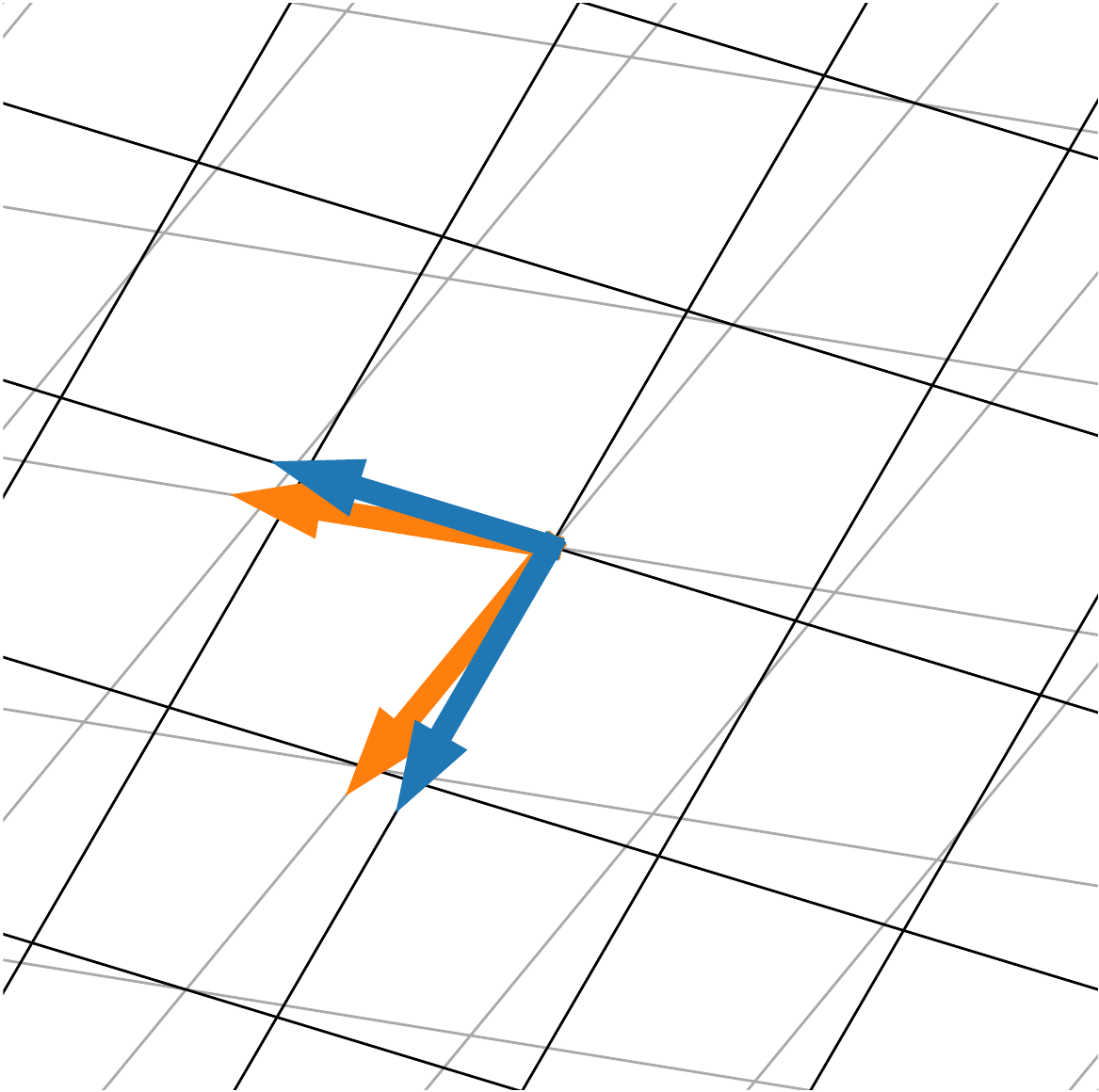}}\\
$\Lambda_A$
& $\Lambda_B$
& Best rigid match\\
\end{tabular}
\caption{(\textbf{Left}) Target lattice, $\Lambda_A$.  The lattice can be described by an infinite number of bases, which differ by a unimodular matrix, \set{L}.  Three such lattice bases, $\set{A}\set{L}_1$, $\set{A}\set{L}_2$, and $\set{A}\set{L}_3$, are shown in orange, green, and red.  
(\textbf{Centre}) A lattice, $\Lambda_B$, whose distance from $\Lambda_A$ we wish to determine.  A single basis, \set{B}, is shown.  
(\textbf{Right}) The correspondence $\set{A}\set{L}_1 = \deformation \set{B}$ has the lowest associated strain.  Shown here is the best rigid match (prior to the stretch operation).
\label{fig:fixed_illustration}}
\end{figure}
}

\subsection{Lattice Correspondences}

The similarity function, $d$, compares two lattices \emph{bases}.  However, every lattice is generated by an infinite number of bases.  Any two bases, $\set{A}$ and $\set{A}^\prime$, which generate the same lattice $\Lambda_A$ are related by $\set{A}^\prime = \set{A} \unimodular$, where $\unimodular \in SL^\pm_3 \left( \vectorspace{Z}{} \right)$ is a is a unimodular matrix, that is $\unimodular \in \vectorspace{Z}{3 \times 3}$ where $\det \left( \unimodular \right) = \pm 1$.  We call $\unimodular$ the \emph{correspondence matrix}.

In order to compare two lattices, rather than specific lattice bases, we optimize the function:
\begin{equation}
\min_{\unimodular \in SL_3 \left( \vectorspace{Z}{} \right)} d \left( \set{A} \unimodular, \set{B} \right)
\label{eq:full_lattice_comparison}
\end{equation}
which compares all possible bases of the first lattice against a single basis of the second, and in doing so finds the optimal lattice correspondence.  This process is illustrated in Figure~\ref{fig:fixed_illustration}.
For any two lattices $\Lambda_A$ and $\Lambda_B$, it is easily shown that the minimum distance shown in Equation~\eqref{eq:full_lattice_comparison} is invariant to the choice of lattice bases.

An efficient algorithm for determining the optimal lattice correspondence is described by Chen \emph{et al.}\cite{chen2016determination}.
Their algorithm proceeds by bounding the maximum norm of the optimal correspondence matrix, and testing all correspondences that lie within that sphere.  We refer to their paper for further details, as it applies to correspondences between fixed lattices only.

\section{Quantification of Symmetry Breaking}
\label{sec:symmetry_breaking}

In this section we extend the concept of the distance function for fixed lattices to variable lattices, and demonstrate its application to the symmetrization of Bravais lattices.
After describing a function for quantification of symmetry breaking, we analyze some properties of the function, and provide some visual examples.

\begin{table}
\begin{tabularx}{\columnwidth}{|lXr|}
\hline
&&\\
\hspace{1mm}\textbf{Parameters:}\hspace{3mm}
& $\set{B} \in \vectorspace{R}{3}$	& \vareqnum\label{var:params}\hspace{1mm}\Bstrut{2}\\
\hspace{1mm}\textbf{Variables:}
& $\set{Z} \in \vectorspace{R}{3}$ & \vareqnum\label{var:var_z}\hspace{1mm}\Bstrut{2}\\
& $\set{G} \in \vectorspace{R}{3}$ \hspace{2mm} $\set{G} = \set{G}^T$ & \vareqnum\label{var:var_gramian}\hspace{1mm}\Bstrut{3}\\
\hspace{1mm}\textbf{Minimize:}
& $3 + \Tr \left( \set{B}^{-T} \set{G} \set{B}^{-1} \right) - 2\Tr \left( \set{Z} \set{B}^{-1} \right)$
& \vareqnum\label{var:objective}\hspace{1mm}\Bstrut{3}\\
\hspace{1mm}\textbf{Subject to:}
& $\begin{bmatrix}
\set{I} & \set{Z}\\
\set{Z}^T & \set{G}\\
\end{bmatrix}
\succeq 0$ & \vareqnum\label{var:schur_gramian}\hspace{1mm}\Bstrut{4}\\
\hline
\end{tabularx}
\caption{Semidefinite program to calculate the minimum-strain symmetrization of a lattice basis, $\set{B}$. 
Here, $\set{Z}$ is a lattice basis matrix whose geometry respects the symmetry conditions of the target Bravais type.  The symmetry conditions are enforced using additional linear constraints on the Gramian matrix, $\set{G}$. 
\label{table:sdp_variable}}
\end{table}

\begin{figure*}
\centering
\begin{tabular}{ m{2.1cm} m{6.5cm} m{2.1cm} m{6cm}}
\multicolumn{2}{l}{\figtext{Primitive cubic}}
& \multicolumn{2}{l}{\figtext{Primitive orthorhombic}}\\
\includegraphics[width=0.09\textwidth]{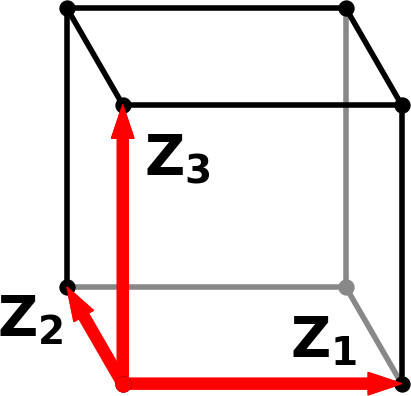}
&
\begin{tabular}{l}
$G_{11} = G_{22} = G_{33}$\Bstrut{2}\\
$G_{12} = G_{13} = G_{23} = 0$\Bstrut{2}\\
\end{tabular}
&
\includegraphics[width=0.09\textwidth]{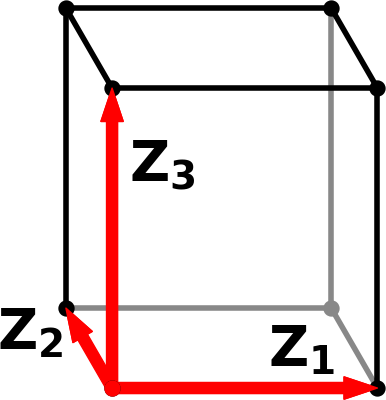}
&
\begin{tabular}{l}
$G_{12} = G_{13} = G_{23} = 0$\Bstrut{2}\\
\end{tabular}
\\\\
\multicolumn{2}{l}{\figtext{Body-centred cubic}}
& \multicolumn{2}{l}{\figtext{Body-centred orthorhombic}}\\
\includegraphics[width=0.09\textwidth]{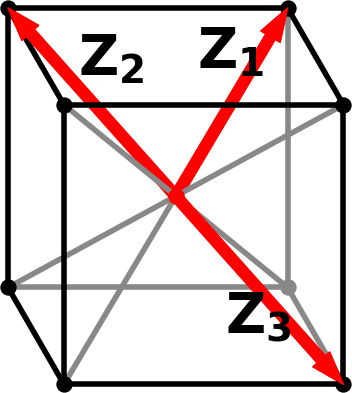}
&
\begin{tabular}{l}
$G_{11} = G_{22} = G_{33}$\Bstrut{2}\\
$G_{12} = G_{13} = G_{23} = -\frac{1}{3} G_{11}$\Bstrut{2}\\
\end{tabular}
&
\includegraphics[width=0.09\textwidth]{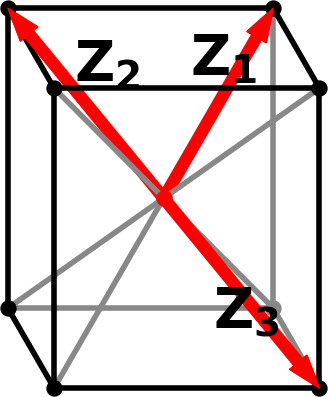}
&
\begin{tabular}{l}
$G_{11} = G_{22} = G_{33}$\Bstrut{2}\\
$G_{12} + G_{13} + G_{23} = -G_{11}$\Bstrut{2}\\
\end{tabular}
\\\\
\multicolumn{2}{l}{\figtext{Face-centred cubic}}
& \multicolumn{2}{l}{\figtext{Base-centred orthorhombic}}\\
\includegraphics[width=0.09\textwidth]{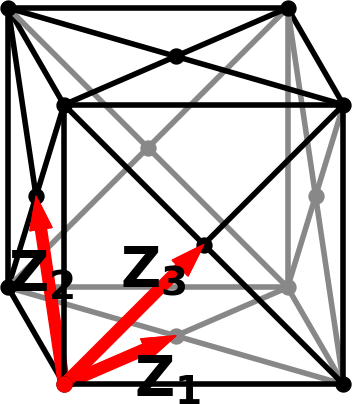}
&
\begin{tabular}{l}
$G_{11} = G_{22} = G_{33}$\Bstrut{2}\\
$G_{12} = G_{13} = G_{23} = \frac{1}{2} G_{11}$\Bstrut{2}\\
\end{tabular}
&
\includegraphics[width=0.09\textwidth]{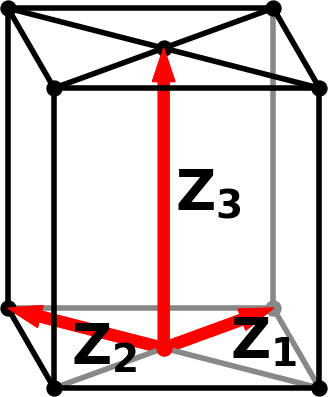}
&
\begin{tabular}{l}
$G_{11} = G_{22}$\Bstrut{2}\\
$G_{13} = G_{23} = 0$\Bstrut{2}\\
\end{tabular}
\\\\
\multicolumn{2}{l}{\figtext{Primitive hexagonal}}
& \multicolumn{2}{l}{\figtext{Face-centred orthorhombic}}\\
\includegraphics[width=0.09\textwidth]{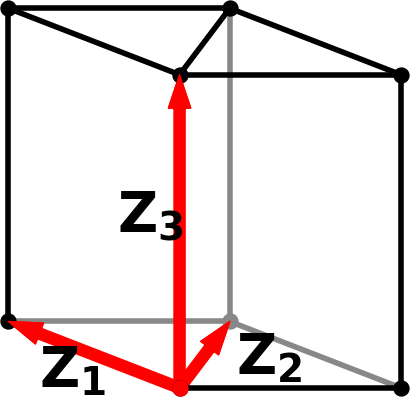}
&
\begin{tabular}{l}
$G_{11} = G_{22}$\Bstrut{2}\\
$G_{13} = G_{23} = 0$\Bstrut{2}\\
$G_{12} = \frac{1}{2} G_{11}$\Bstrut{2}\\
\end{tabular}
&
\includegraphics[width=0.09\textwidth]{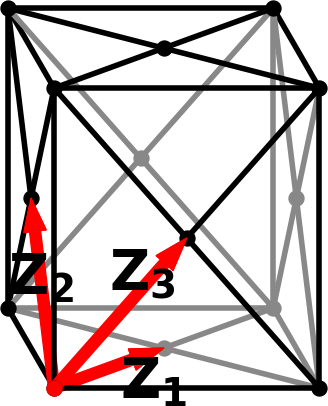}
&
\begin{tabular}{l}
$G_{11} = G_{12} + G_{13}$\Bstrut{2}\\
$G_{22} = G_{12} + G_{23}$\Bstrut{2}\\
$G_{33} = G_{13} + G_{23}$\Bstrut{2}\\
\end{tabular}
\\\\
\multicolumn{2}{l}{\figtext{Primitive rhombohedral}}
& \multicolumn{2}{l}{\figtext{Primitive monoclinic}}\\
\includegraphics[width=0.09\textwidth]{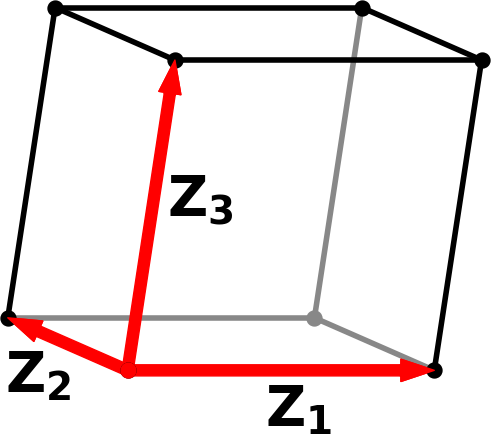}
&
\begin{tabular}{l}
$G_{11} = G_{22} = G_{33}$\Bstrut{2}\\
$G_{12} = G_{13} = G_{23}$\Bstrut{2}\\
\end{tabular}
&
\includegraphics[width=0.09\textwidth]{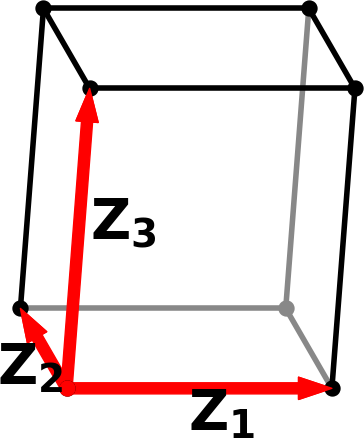}
&
\begin{tabular}{l}
$G_{12} = G_{23} = 0$\Bstrut{2}\\
\end{tabular}
\\\\
\multicolumn{2}{l}{\figtext{Primitive tetragonal}}
& \multicolumn{2}{l}{\figtext{Base-centred monoclinic}}\\
\includegraphics[width=0.09\textwidth]{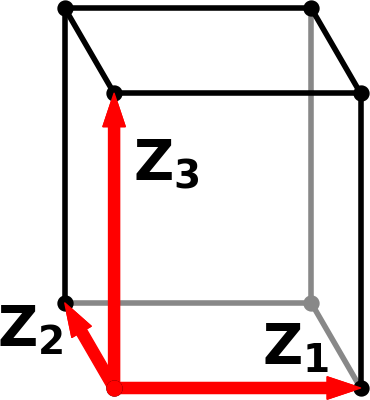}
&
\begin{tabular}{l}
$G_{11} = G_{22}$\Bstrut{2}\\
$G_{12} = G_{13} = G_{23} = 0$\Bstrut{2}\\
\end{tabular}
&
\includegraphics[width=0.09\textwidth]{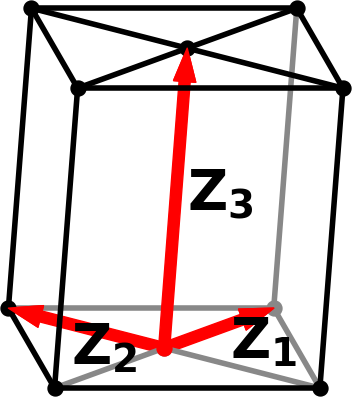}
&
\begin{tabular}{l}
$G_{11} = G_{22}$\Bstrut{2}\\
$G_{13} = G_{23} = 0$\Bstrut{2}\\
\end{tabular}
\\\\
\multicolumn{2}{l}{\figtext{Body-centred tetragonal}}
& \multicolumn{2}{l}{\figtext{Primitive triclinic}}\\
\includegraphics[width=0.09\textwidth]{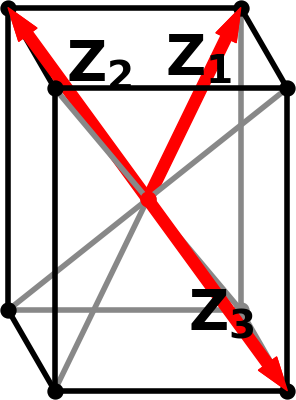}
&
\begin{tabular}{l}
$G_{11} = G_{22} = G_{33}$\Bstrut{2}\\
$G_{12} + G_{13} + G_{23} = -G_{11}$\Bstrut{2}\\
$G_{13} = G_{23}$\Bstrut{2}\\
\end{tabular}
&
\includegraphics[width=0.09\textwidth]{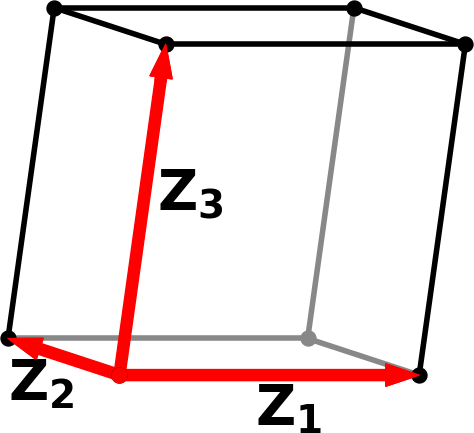}
&
\end{tabular}
\caption{Linear constraints on the Gramian matrices for each of the fourteen three-dimensional Bravais lattices.  The constraints are constructed such that the symmetry conditions of the Bravais lattice are respected.
The triclinic lattice type does not have any constraints on the form of the Gramian; all lattices are trivially triclinic, since there are no symmetry conditions to satisfy.
\label{fig:lattice_gramians}}
\end{figure*}

\subsection{Variable Lattice Bases}

The distance function in Equation~(\ref{eq:distance_strain}) compares two lattices with fixed parameters.  In order to quantify symmetry breaking, we introduce lattices with \emph{variable} lattice parameters.  As stated above, the problem we aim to solve is the following: given an observed lattice which we which to symmetrize, find the closest lattice in the target Bravais class.  We will first consider this problem for a fixed correspondence.

Let $\set{B} \in \vectorspace{R}{3 \times 3}$ be a lattice basis which we wish to symmetrize, and let $\set{Z} \in \vectorspace{R}{3 \times 3}$ be a variable lattice basis.
The symmetrized basis is found by solving the problem
\begin{equation}
\set{Z}^* = \argmin_{\set{Z} \in \vectorspace{R}{3 \times 3}} d_B \left( \set{Z}, \set{B} \right)
\label{eq:optimal_z}
\end{equation}
Here, $d_B$ denotes the distance function in Equation~\eqref{eq:distance_strain}
subject to geometric constraints on $\set{Z}$ which enforce the symmetry conditions of a chosen Bravais lattice type ($B$) (described below).

We can rearrange the distance function in terms of the Frobenius and nuclear norms of $\set{Z}\set{B}^{-1}$.  For a matrix $\set{X} \in \vectorspace{R}{3 \times 3}$ with singular values $\left[ \sigma_1, \sigma_2, \sigma_3 \right]$, the Frobenius norm is $\frobnorm{\set{X}} = \sqrt{\sigma_1^2 + \sigma_2^2 + \sigma_3^2}$ and the nuclear norm is $\nuclearnorm{\set{X}} = \sigma_1 + \sigma_2 + \sigma_3$.
The distance function is then
\begin{equation}
\begin{split}
d \left( \set{Z}, \set{B} \right)
&= \sqrt{ 3 + \frobnorm{\set{Z}\set{B}^{-1}}^2 - 2 \nuclearnorm{\set{Z}\set{B}^{-1}} }\\
\end{split}
\label{eq:distance_trace_norm}
\end{equation}
Let $\set{G} = \set{Z}^T \set{Z}$ be the Gramian
\footnote{For Bravais lattice bases the Gramian and the metric tensor are equivalent.  The former term is used in matrix algebra.} 
matrix of $\set{Z}$.
Then, $\Tr \left( \set{B}^{-T} \set{G} \set{B}^{-1} \right)= \frobnorm{ \set{Z}\set{B}^{-1} }^2$.
Using the Gramian form, the minimization problem in Equation~\eqref{eq:optimal_z} can be expressed as a \emph{semidefinite program} (SDP).  Semidefinite programming is a form of convex optimization which generalizes linear programming~\cite{vandenberghe1996semidefinite}; in addition to a linear objective and linear constraints, a SDP permits positive semidefiniteness constraints on matrix variables.

Table~\ref{table:sdp_variable} shows a SDP for solving Equation~\eqref{eq:optimal_z}.
The input parameter is the lattice basis we wish to symmetrize \vareqref{var:params}.  The matrix variables are the variable lattice basis \vareqref{var:var_z} and the Gramian matrix \vareqref{var:var_gramian}.  The objective is to minimize the square of the distance function \vareqref{var:objective}.
By the Schur complement condition for positive semi-definiteness~\cite{zhang2006schur,boyd2004convex}, the linear matrix inequality in \vareqref{var:schur_gramian} enforces the condition $\set{G} - \set{Z}^T \set{Z} \succeq 0$ (where the relation $\set{X} \succeq 0$ means $\set{X}$ is positive semidefinite).
At the optimum value of \set{Z} this inequality is tight, i.e~$\set{G} = \set{Z}^T \set{Z}$.
In addition to the constraint \vareqref{var:schur_gramian}, we impose a set of linear constraints on $\set{G}$ in order to control the geometry of $\set{Z}$ and thereby enforce the symmetry conditions of a chosen Bravais lattice type.  The constraints for each Bravais type are shown in Figure~\ref{fig:lattice_gramians}.

The SDP shown above finds a symmetrized basis, $\set{Z}^*$, by simultaneously finding the lattice parameters and the lattice rotation.  This solution is the closest lattice basis to $\set{B}$ under the defined distance function.  The basis respects the specified symmetry constraints, and represents the minimum-strain symmetrization of the basis $\set{B}$.
The resulting distance quantifies the symmetry breaking of the target Bravais type.

\begin{figure}
\begin{algorithm}[H]
\begin{algorithmic}[1]
\Procedure{Symmetrize}{$\set{B}$, $B$, $\mathcal{L}$}
	\State $d^* = \infty$				\Comment{Optimal distance} \label{alg:best_distance}
	\State $\unimodular^* = \set{I}$	\Comment{Optimal correspondence} \label{alg:best_correspondence}
	\State visited $:=$ $\emptyset$		\Comment{`Visited' correspondences} \label{alg:visited}
	\State found $:=$ True
	\While{found}	\label{alg:iteration}
		\State found $:=$ False
		\State $\unimodular_c = \unimodular^*$

		\For {$\unimodular \in \mathcal{L}$} \label{alg:neighbourhood}
			\If {$\unimodular_c \unimodular \notin$ visited}
				\State $\set{Z}^* := \argmin_{\set{Z} \in \vectorspace{R}{3 \times 3}} d_B \left( \set{Z} \unimodular_c \unimodular, \set{B} \right)$
				\If {$d_B \left( \set{Z}^* \unimodular_c \unimodular, \set{B} \right) < d^*$} \label{alg:distance_test}
					\State $d^* := d_B \left( \set{Z}^* \unimodular_c \unimodular, \set{B} \right)$ \label{alg:distance_update}
					\State $\unimodular^* := \unimodular_c \unimodular$ \label{alg:correspondence_update}
					\State found $:=$ True
				\EndIf
			\EndIf
			\State visited $:=$ visited $\cup \{ \unimodular_c \unimodular \}$
		\EndFor
	\EndWhile
	\State \Return $\{ d^*, \unimodular^* \}$
\EndProcedure
\end{algorithmic}
\caption{A steepest-descent algorithm for finding the optimal lattice correspondence.
\label{alg:correspondence}}
\end{algorithm}
\end{figure}

\subsection{Lattice Correspondence Search}

By formulating the distance function as a SDP, we have shown that the distance function is convex for a fixed lattice correspondence.  In order to find the minimum-strain solution, however, we need to optimize over all lattice correspondences, of which there are an infinite number.  For fixed lattices, the optimal correspondence can be found efficiently, owing to the existence of an upper bound on the norm of the optimal correspondence matrix~\cite{chen2016determination}.  For variable lattices no such bound is known.  Instead, we use a steepest descent search over lattice correspondences.

Starting from a good candidate correspondence, we search the surrounding correspondences, iteratively, until no better solution can be found.  For the initial `good candidate' correspondence, we use a right-handed Minkowski-reduced~\cite{minkowski1883reduction} (MR) basis of the $\Lambda_B$.  A MR basis is a lattice basis whose lattice vectors are the shortest possible.  It can be computed efficiently~\cite{nguyen2009revisited}.

We define the neighbourhood of a correspondence as all unimodular matrices with positive determinant and whose elements have magnitude at most 1:
\begin{equation}
\mathcal{L} = \left\{ \unimodular \in SL^+_3 \left( \vectorspace{Z}{} \right) \; \mid \; -1 \leq l_{ij} \leq 1  \;\; \forall i, j\right\}
\label{eq:correspondence_neighbourhood}
\end{equation}

The steepest-descent approach is shown in Algorithm~\ref{alg:correspondence}.
The inputs to the algorithm are: a MR-basis ($\set{B}$) of the lattice we wish to symmetrize, the target Bravais type ($B$), and the correspondence neighbourhood ($\mathcal{L}$).  The algorithm keeps track of the minimum distance (line~\ref{alg:best_distance}) and the optimal correspondence (line~\ref{alg:best_correspondence}), and maintains a set of visited correspondences (line~\ref{alg:visited}) i.e.~correspondences which have been tested.
In each iteration (line~\ref{alg:iteration}) the search tests all correspondences in the neighbourhood of the current best solution (line~\ref{alg:neighbourhood}).  Each time a correspondence is tested (line~\ref{alg:distance_test}), the best solution (lines~\ref{alg:distance_update}-\ref{alg:correspondence_update}) are updated as needed.  Here $d_B$ denotes the distance function with the appropriate linear constraints on $\set{G}$ for a chosen Bravais type, $B$.
The algorithm iteratively tests the neighbourhood of the current best solution until no better solution can be found.

\begin{figure}
\centering
{
\setlength{\fboxsep}{0pt}%
\setlength{\fboxrule}{1pt}%
\fbox{\includegraphics[trim={0 1.3cm 0 1.3cm},clip, width=0.75\columnwidth]{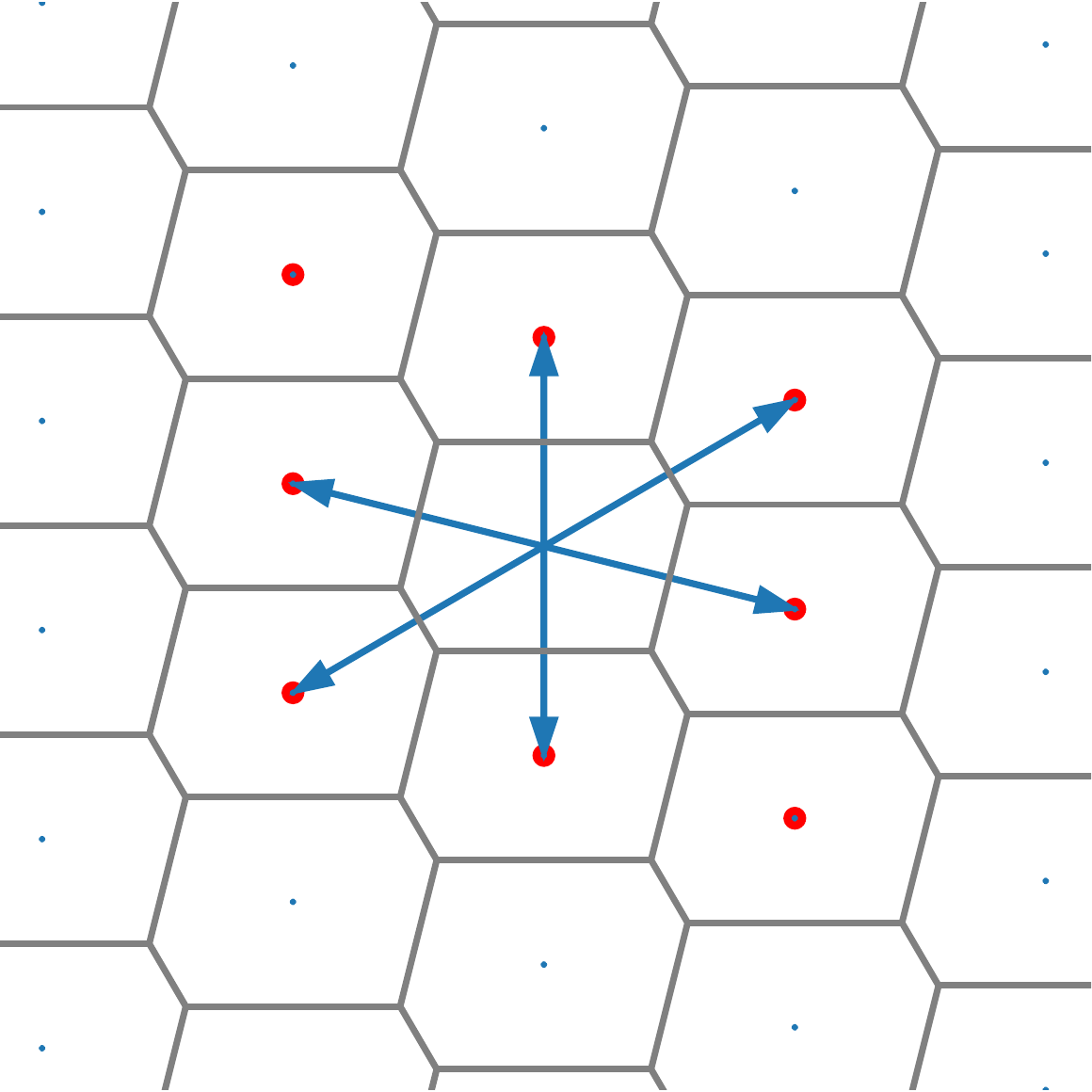}}
}
\caption{Voronoi cells (Wigner-Seitz cells) in a 2D oblique lattice.  Voronoi-relevant vectors are those which pass through a Voronoi face to an adjacent lattice point; they constitute a subset of the red lattice points, which are combinations of $\left\{-1, 0, 1 \right\}$ of the MR basis vectors.
\label{fig:voronoi_relevant}}
\end{figure}

The neighbourhood we define in Equation~\ref{eq:correspondence_neighbourhood} is motivated by the need for a distance function which is continuous in the presence of a continuous lattice deformation.  The distance function is continuous if we consider, at a minimum, all lattice bases consisting of Voronoi-relevant vectors~\cite{micciancio2013voronoi}.  Voronoi-relevant vectors are illustrated in Figure~\ref{fig:voronoi_relevant}.
A theorem due to Minkowski~\cite{minkowski1905voronoirelevant} states that the Voronoi-relevant vectors are a non-strict subset of all integer combinations of $\{-1, 0, 1\}$ of the MR basis vectors.  The neighbourhood defined in Equation~\eqref{eq:correspondence_neighbourhood} contains all right-handed unimodular matrices of this form.

\begin{figure}
\centering
\begin{tabular}{ccc}
\includegraphics[width=0.32\columnwidth]{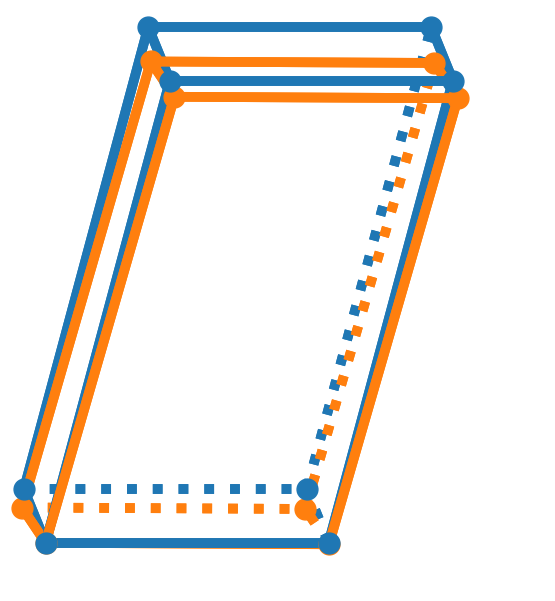}
& \includegraphics[width=0.32\columnwidth]{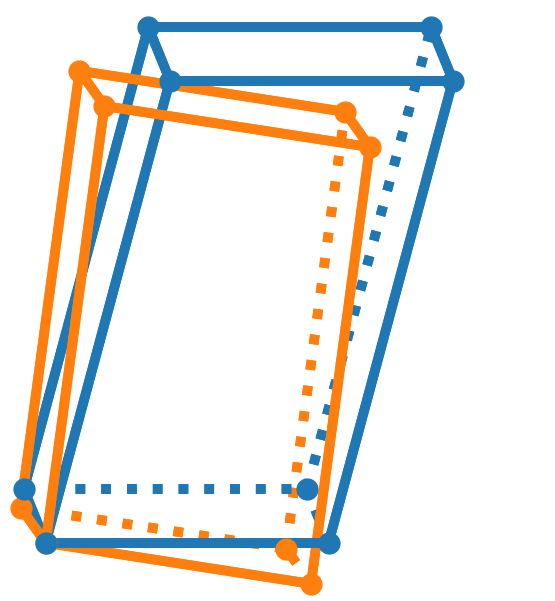}
& \includegraphics[width=0.32\columnwidth]{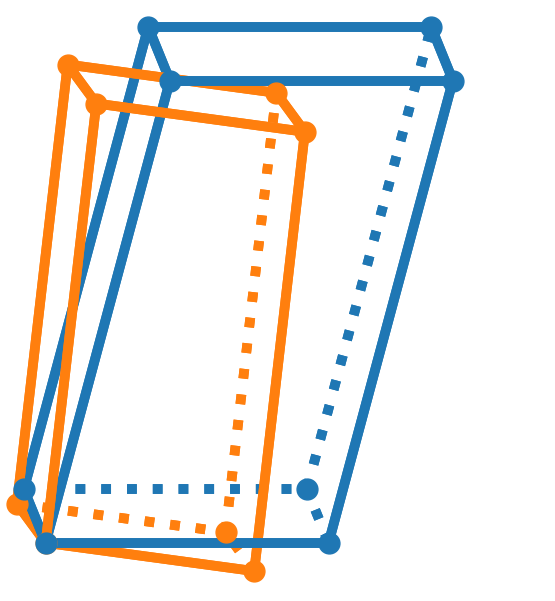}\\
\figtext{Monoclinic} & \figtext{Orthorhombic} & \figtext{Tetragonal}\\
$\frobnorm{\set{E}} = 0.131$
& $\frobnorm{\set{E}} = 0.246$
& $\frobnorm{\set{E}} = 0.355$\\\\
\includegraphics[width=0.32\columnwidth]{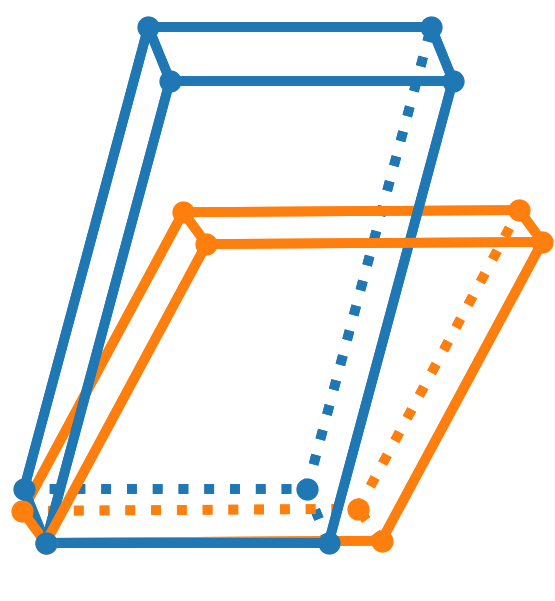}
& \includegraphics[width=0.32\columnwidth]{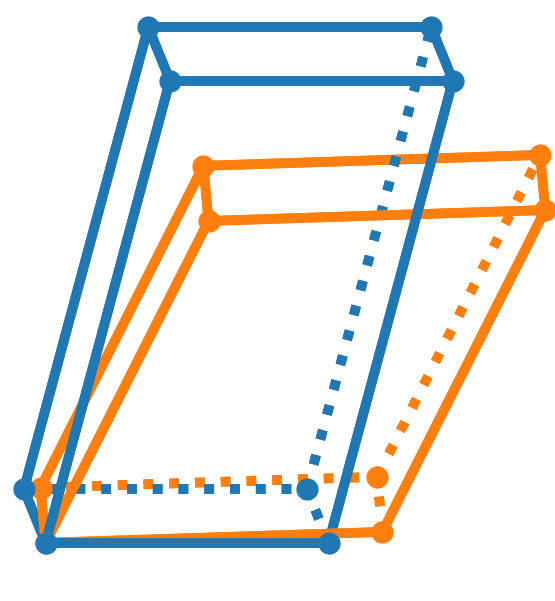}
& \includegraphics[width=0.32\columnwidth]{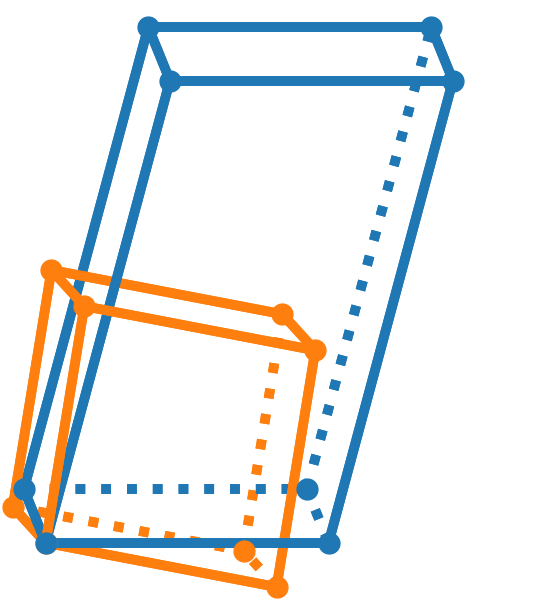}\\
\figtext{Hexagonal} & \figtext{Rhombohedral} & \figtext{Cubic}\\
$\frobnorm{\set{E}} = 0.404$
& $\frobnorm{\set{E}} = 0.412$
& $\frobnorm{\set{E}} = 0.593$
\end{tabular}
\caption{Successively aggressive lattice symmetrizations.  A triclinic cell (blue) is symmetrized to each of the six primitive Bravais lattice types.  The symmetrized cells (orange) can be mapped onto the triclinic cell with a pure stretch.  The lattice parameters and rotations of the symmetrized cells represent minimum-strain solutions.
\label{fig:symmetrization_example}}
\end{figure}


With the lattice correspondence search in place, we can perform minimum-strain symmetrization.  This is illustrated for a selection of lattice types in Figure~\ref{fig:symmetrization_example}.  Of the lattices shown, only the monoclinic symmetrization has a small strain tensor norm.  Nonetheless, even for very dissimilar lattices types with a correspondingly large strain tensor, we can still compute a minimum-strain symmetrization.

\subsection{Scale Invariance and Range of the Symmetrization Distance}

We have described minimum-strain symmetrization for the different Bravais lattice types.  Here, we show that the distance function is scale invariant, and that the 
scale invariance sets an upper bound on the range of the distance function.

When comparing two fixed lattices, $d \left( \set{A}, \set{B} \right)$ is dependent on the scales of \set{A} and \set{B}.  However, when using variable lattices to quantify symmetry breaking, the resulting distance is scale invariant.  For a scaling factor, $k \in \vectorspace{R}{}$, we have
\begin{equation}
\argmin_{\set{Z} \in \vectorspace{R}{3 \times 3}} d_B \left( \set{Z}, \set{B} \right)
= \argmin_{\set{Z} \in \vectorspace{R}{3 \times 3}} d_B \left( \set{Z}, k \set{B} \right)
\end{equation}
Due to the lattice parameters of $\set{Z}$ being optimized, any change in scale of \set{B} is accommodated by a corresponding change in $\set{Z}$.  This is a particularly useful property when comparing symmetry breaking in cells of different sizes.

The scale invariance also sets an upper bound on the maximum strain.
For a chosen Bravais type and a correspondence matrix $\unimodular$, the range of the distance function $d_B \left( \set{Z} \unimodular, \set{B} \right)$ is $\left[ 0, \sqrt{2} \right)$.
The zero-strain solution is trivial.
To show that the interval is bounded from above by $\sqrt{2}$, we observe that even the most constrained Bravais types (the cubic lattices) accommodate a change in scale of $\set{Z}$.
Representing this by a parameter $s \in \vectorspace{R}{}$, the distance function (minimized over all lattice correspondences) is given by:
\begin{equation}
\begin{split}
&\min_{ \substack{ s \in \vectorspace{R}{} \\  \unimodular \in \slthree}}
d \left( s \set{A} \unimodular, \set{B} \right)\\
=
&\min_{ \substack{ s \in \vectorspace{R}{} \\  \unimodular \in \slthree}}
\frobnorm{ s \sqrt{ \set{B}^{-T} \unimodular^T \set{A}^T \set{A} \unimodular \set{B}^{-1} } - \set{I} }
\end{split}
\end{equation}
Let $\vec{\sigma} = [ \sigma_1, \sigma_2, \sigma_3 ]$ be the singular values of $\set{A}\unimodular\set{B}^{-1}$.  It can be shown that the optimal scaling parameter is given by
\begin{equation}
s^* = \frac{ \sigma_1 + \sigma_2 + \sigma_3 }{ \sigma_1^2 + \sigma_2^2 + \sigma_3^2 }
\end{equation}
Expressed in terms of singular values, the distance function is:
\begin{equation}
\min_{ \substack{ s \in \vectorspace{R}{} \\  \unimodular \in \slthree}}
d \left( s \set{A} \unimodular, \set{B} \right)
= \sqrt{3 - \frac{\left( \sigma_1 + \sigma_2 + \sigma_3 \right)^2}{ \sigma_1^2 + \sigma_2^2 + \sigma_3^2}}
\end{equation}
which attains a maximum (at $\sqrt{2}$) when $\vec{\sigma} = [ \sigma_1, 0, 0 ]$.
The rank of a matrix is equal to the number of non-zero singular values.
Let $\set{B}^{-1}$ be a rank-1 matrix.  Then $\set{A}\unimodular \set{B}^{-1}$ is a rank-1 matrix for every $\unimodular \in \slthree$ and any matrix $\set{A}$, and $\vec{\sigma} = [ \sigma_1, 0, 0 ]$.
Since a rank-1 matrix is not invertible, this vector of singular values can only be attained asymptotically.  Let $\set{B} = \diag \left( 1, \kappa, \kappa \right)$. 
Then $\lim_{\kappa \to \infty} \vec{\sigma} = [ 1, 0, 0 ]$.  It follows that
\begin{equation}
\lim_{\kappa \to \infty} \min_{ \substack{ s \in \vectorspace{R}{} \\ \unimodular \in \slthree}}
d \left( s \set{A} \unimodular, \diag\left( 1, \kappa, \kappa \right) \right) = \sqrt{2}
\end{equation}

The analysis shown here establishes the upper bound of the distance function.  A similar analysis for Bravais types with more degrees of freedom is significantly more complicated.  The additional degrees of freedom, however, can only serve to reduce the range of the distance function.

\newcolumntype{L}[1]{>{\raggedright\arraybackslash}m{#1}}  
\newcolumntype{C}[1]{>{\centering\arraybackslash}m{#1}}   

\begin{figure*}
\centering
\begin{tabular}{C{0.78\textwidth}L{0.22\textwidth}}
\figtext{Primitive tetragonal}
& \figtext{Cell parametrization}\\
\includegraphics[width=0.76\textwidth]{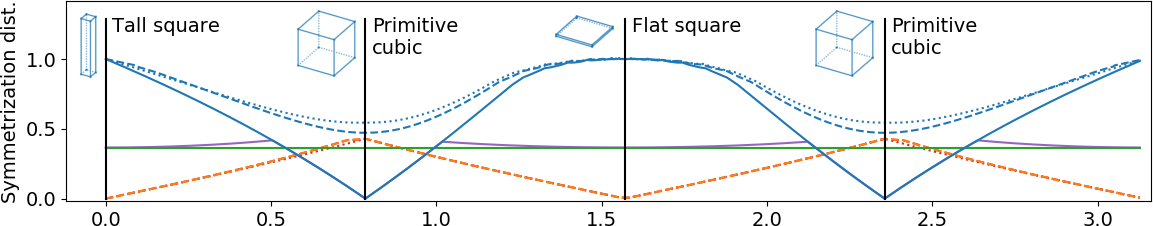}
&
$\displaystyle
\begin{bmatrix}
\sin \phi & 0 & 0\\
0 & \sin \phi & 0\\
0 & 0 & \cos \phi\\
\end{bmatrix}$
\vspace{2mm}\\
{$\bm{\phi}$}\vspace{2mm}\\
\figtext{Body centred tetragonal}
& \figtext{Cell parametrization}\\
\includegraphics[width=0.76\textwidth]{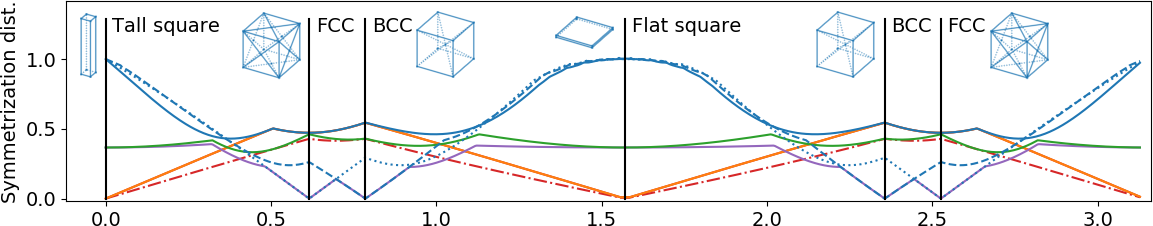}
&
$\displaystyle
\begin{bmatrix}
-\sin \phi & \sin \phi & \sin \phi\\
\sin \phi & -\sin \phi & \sin \phi\\
\cos \phi & \cos \phi & -\cos \phi\\
\end{bmatrix}$
\vspace{2mm}\\
{$\bm{\phi}$}\vspace{3mm}\\
\includegraphics[width=0.05\textwidth]{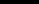} \figtext{Primitive}\hspace{3mm}
\includegraphics[width=0.05\textwidth]{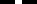} \figtext{Body-centred}\hspace{3mm}
\includegraphics[width=0.05\textwidth]{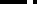} \figtext{Base-centred}\hspace{3mm}
\includegraphics[width=0.05\textwidth]{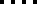} \figtext{Face-centred}
\vspace{1mm}\\
\includegraphics[width=0.016\textwidth]{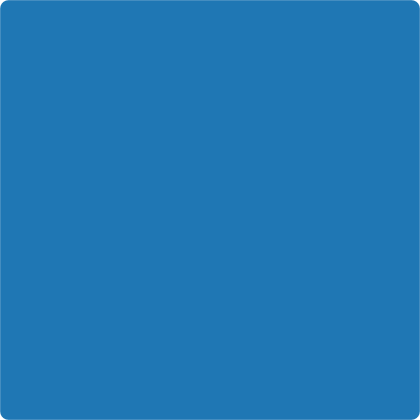}
\figtext{Cubic}\hspace{3mm}
\includegraphics[width=0.016\textwidth]{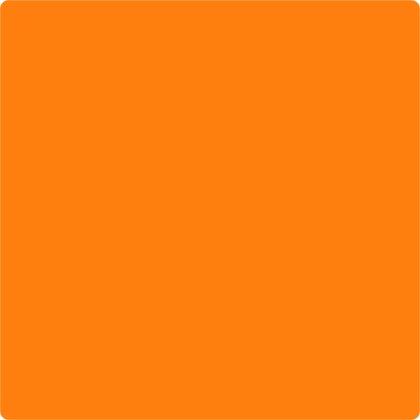}
\figtext{Tetragonal}\hspace{3mm}
\includegraphics[width=0.016\textwidth]{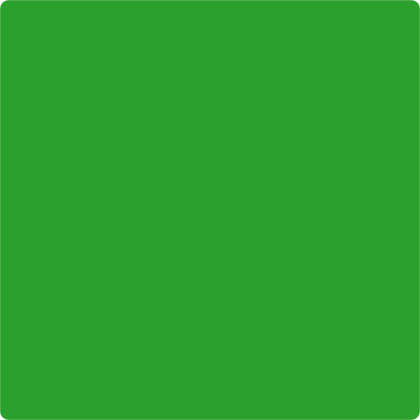}
\figtext{Hexagonal}\hspace{3mm}
\includegraphics[width=0.016\textwidth]{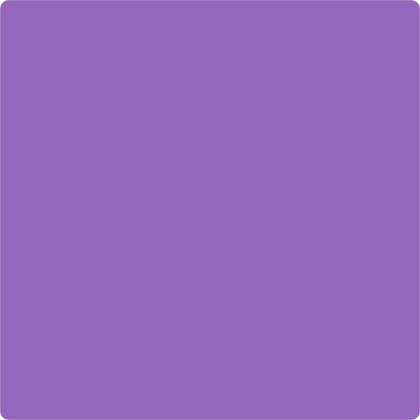}
\figtext{Rhombohedral}\hspace{3mm}
\includegraphics[width=0.016\textwidth]{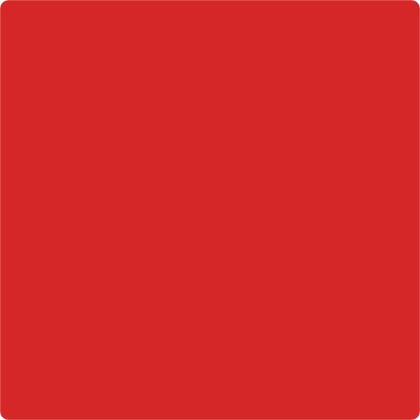}
\figtext{Orthorhombic}
\end{tabular}
\caption{Symmetrization distances for all geometries of the primitive and body-centred tetragonal lattices (modulo scale).
Since the symmetrization distance is scale invariant, these lattices can be parametrized using a single parameter: using the lattice parameters of the conventional setting $a = \sin\phi$ and $c = \cos\phi$.
  Zero-distance symmetrizations are not shown.  In the primitive tetragonal figure, the body-centred orthorhombic symmetrization distances are identical to the body-centred tetragonal distances and are not shown.
  Similarly, in the body-centred tetragonal figure, the primitive monoclinic symmetrization distances are identical to the base-centred monoclinic distances, and the primitive orthorhombic symmetrization distances are identical to the primitive tetragonal distances.
\label{fig:variation_1D}}
\end{figure*}

\subsection{Implementation}

We have presented the symmetrization procedure as a semidefinite program.  In addition to being the most compact description, formulation as a semidefinite program is the most convenient for analysis of the scale invariance and range of the distance function.  For a practical implementation, however, it is less than ideal to rely upon semidefinite programming, as SDP solvers are slow.  
Fortunately, any optimization approach which achieves the same result as the SDP is equally valid.

By employing a quaternion parametrization, symmetrization can be done by solution of a constrained multivariate polynomial, which can be solved efficiently using sequential quadratic programming~\cite{boggs1995sqp}. Further details are given in the appendix.
Compared to the existing Bravais classification methods this procedure has significantly higher computational requirements, taking on the order of $0.2\text{s}$ per cell.  Despite the increased computation time, the relevant comparison is not between classification methods, but to the time required to perform an experiment or an \emph{ab initio} structure calculation.  In this regard, the increased computation time is of little importance.

\begin{figure}[b]
\centering
\includegraphics[width=0.9\columnwidth]{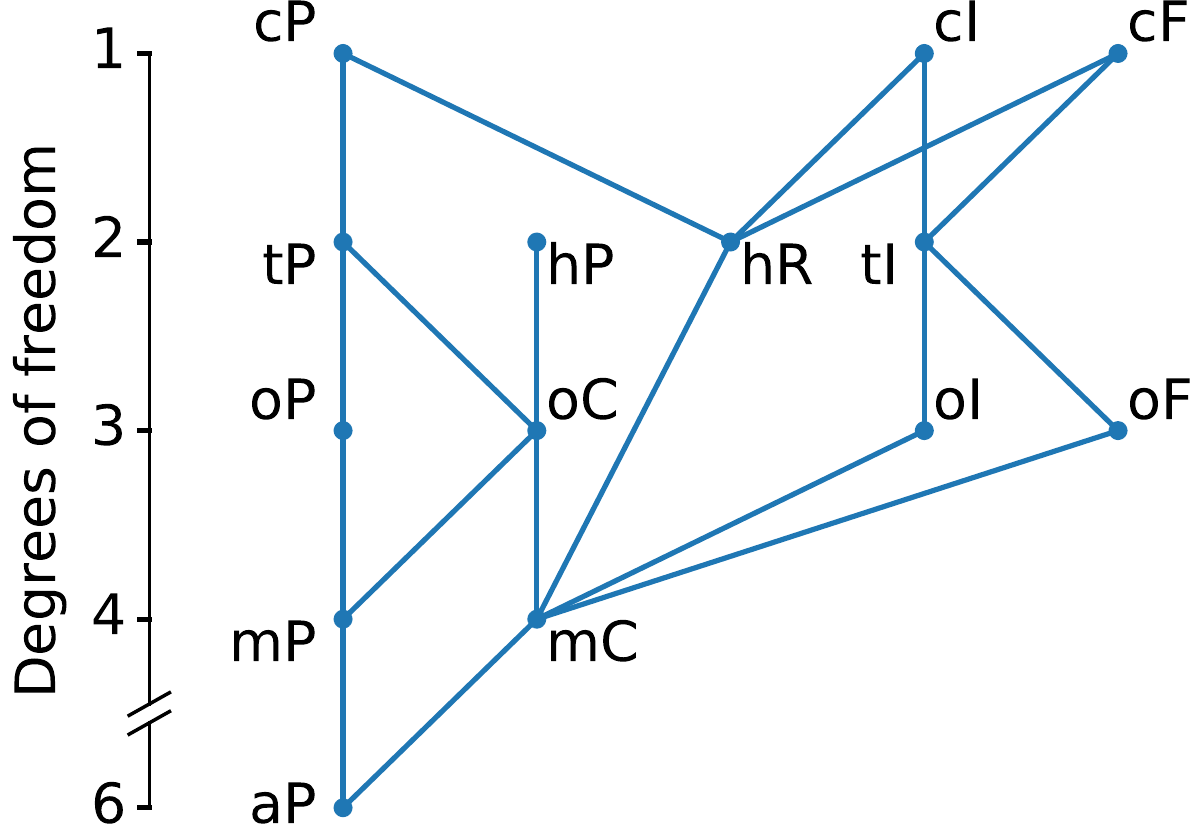}
\caption{Hierarchy of symmetry relationships between Bravais types.  The Bravais types are arranged vertically according to the number of degrees of freedom in the lattice.  Presence of an upwards path indicates a subset relationship.  For clarity, Bravais types are denoted by their Pearson symbols (a: triclinic, m: monoclinic, o: orthorhombic, t: tetragonal, h: hexagonal, c: cubic, P: primitive, C: base-centred, R: rhombohedral, I: body-centred, F: face-centred).
\label{fig:hierarchy}}
\end{figure}

\begin{figure*}
\centering
\begin{tabular}{cc}
\includegraphics[width=0.9\columnwidth]{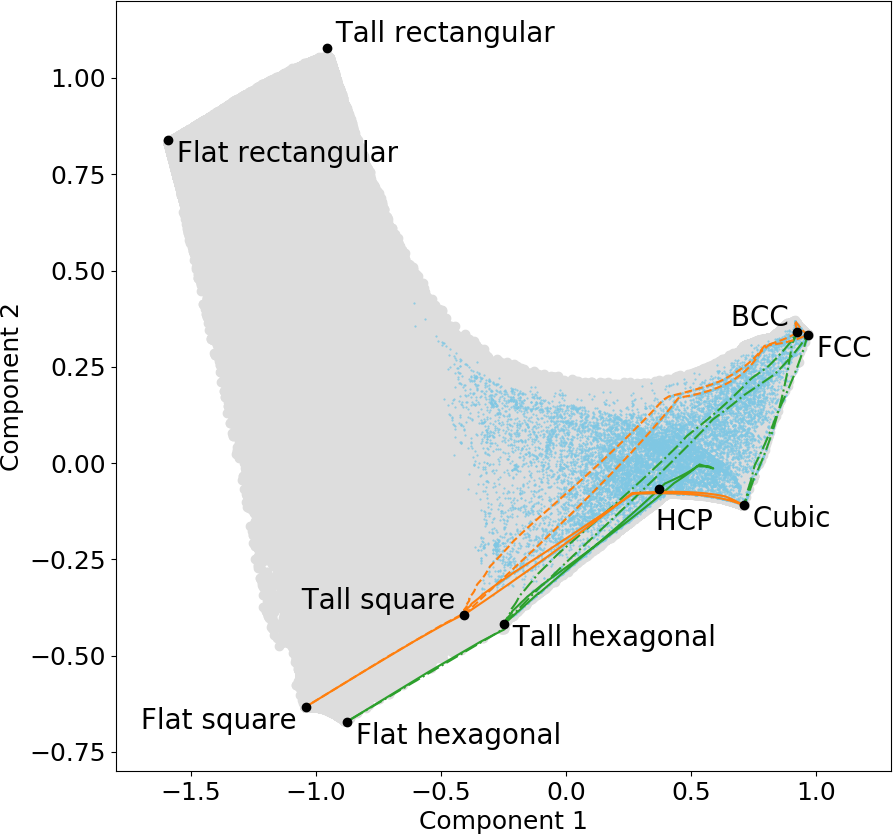}
\hspace{2mm}
&
\hspace{2mm}
\includegraphics[width=0.9\columnwidth]{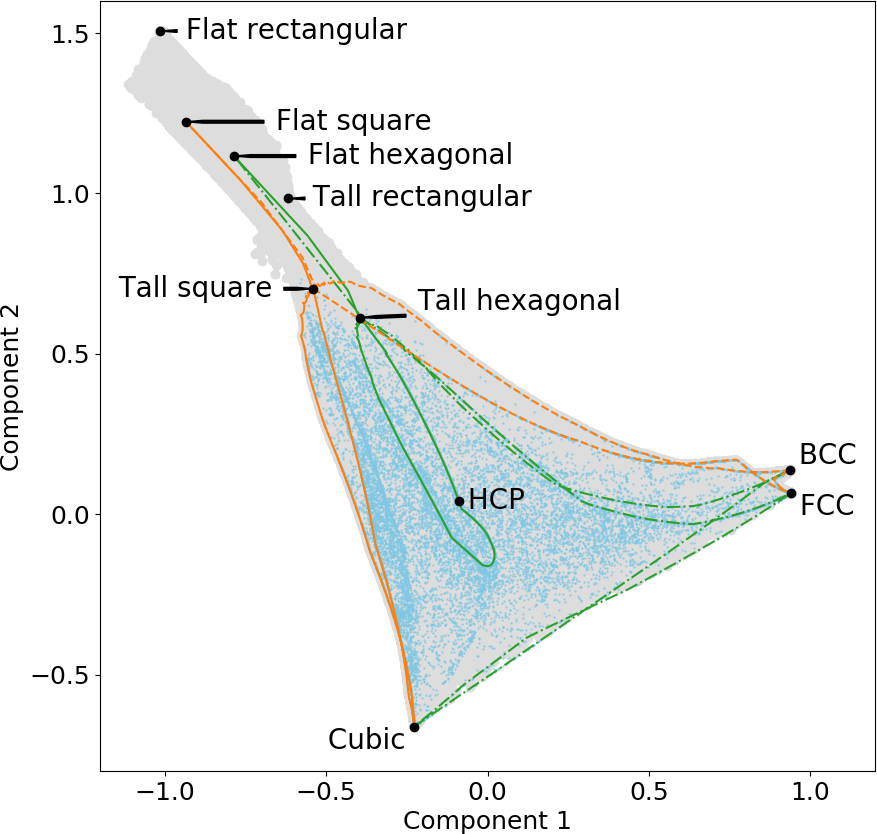}\\
\figtext{(a) Full Bravais-space projection}
& \figtext{(b) COD projection}
\end{tabular}
\vspace{4mm}\\
\begin{tabular}{ccc}
\includegraphics[width=0.6\columnwidth]{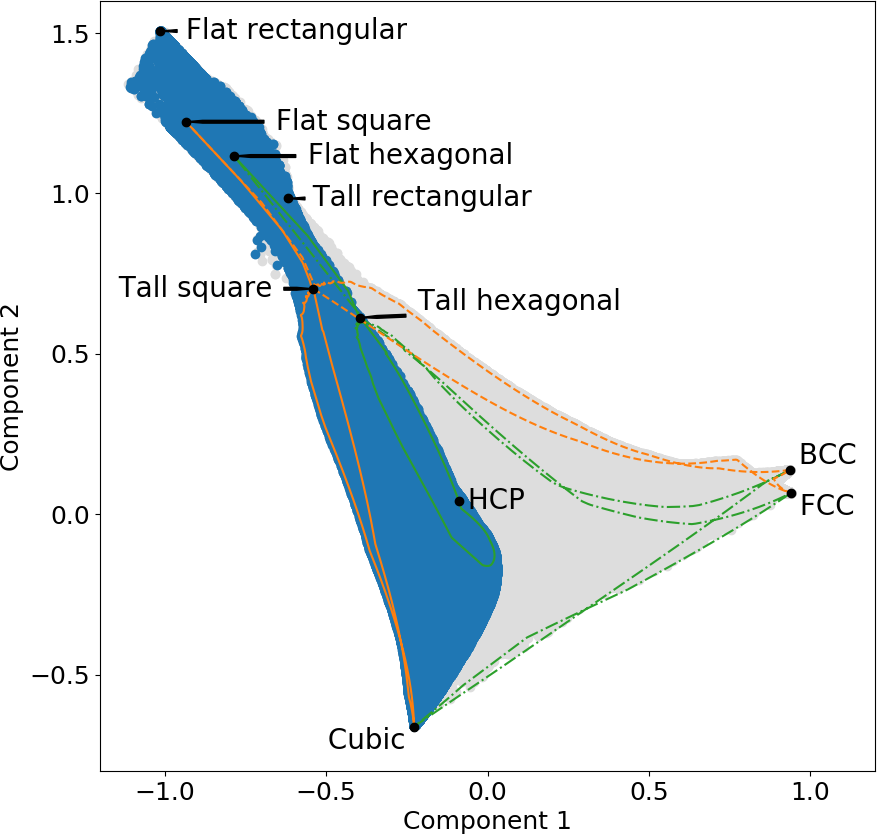}
&\includegraphics[width=0.6\columnwidth]{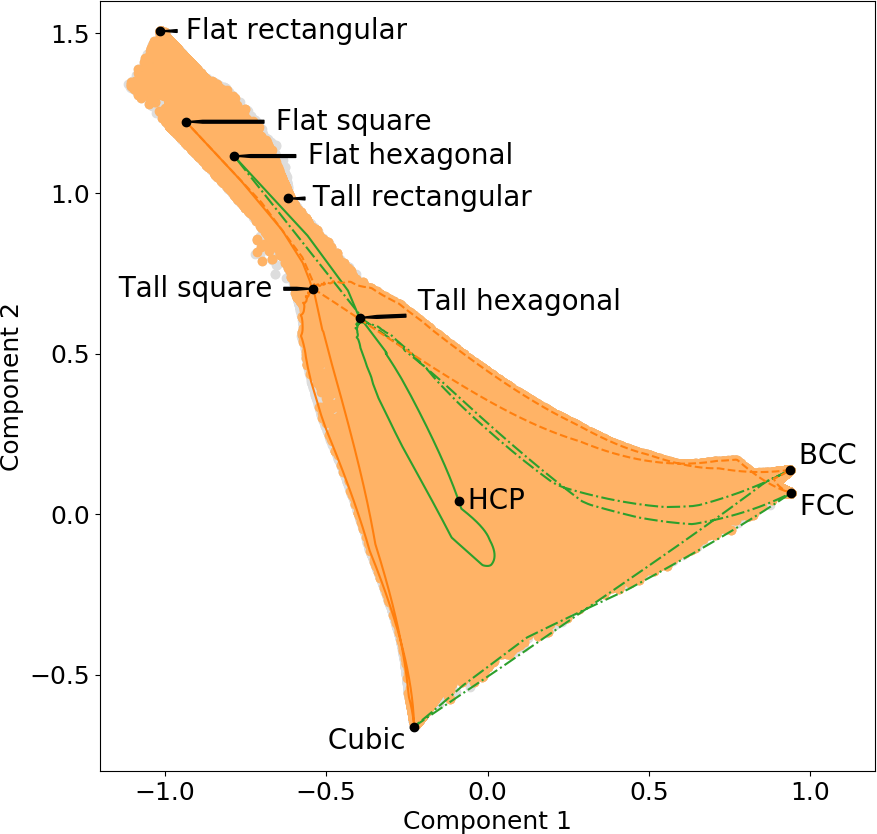}
&\includegraphics[width=0.6\columnwidth]{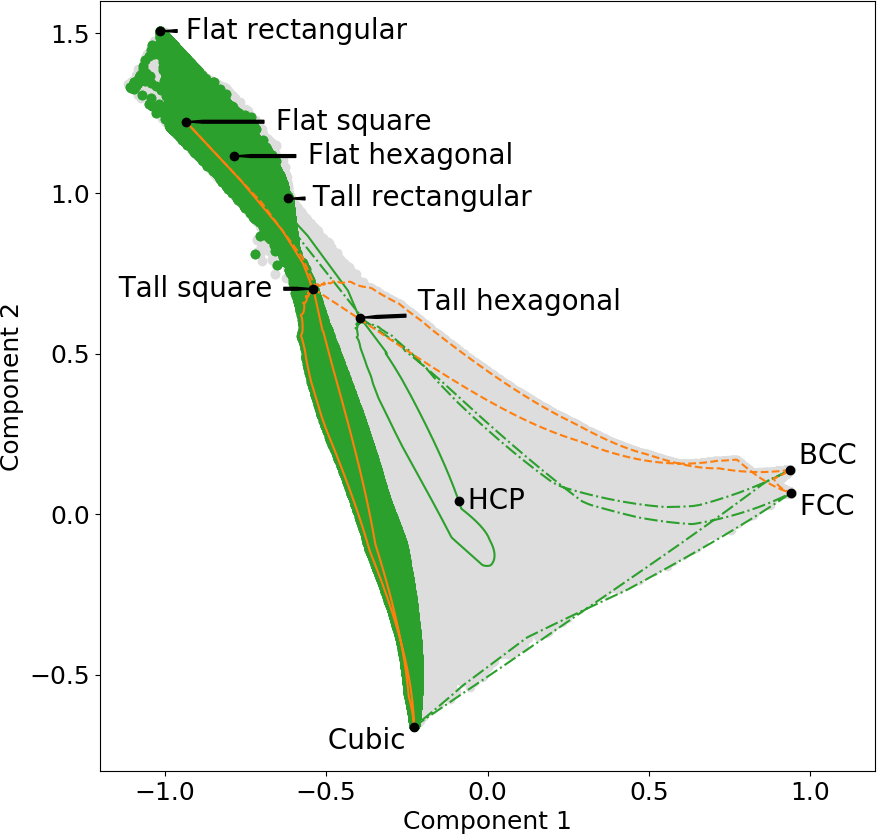}\\
\figtext{(c) Primitive monoclinic}
& \figtext{(d) Base-centred monoclinic}
& \figtext{(e) Primitive orthorhombic}
\vspace{4mm}\\
\includegraphics[width=0.6\columnwidth]{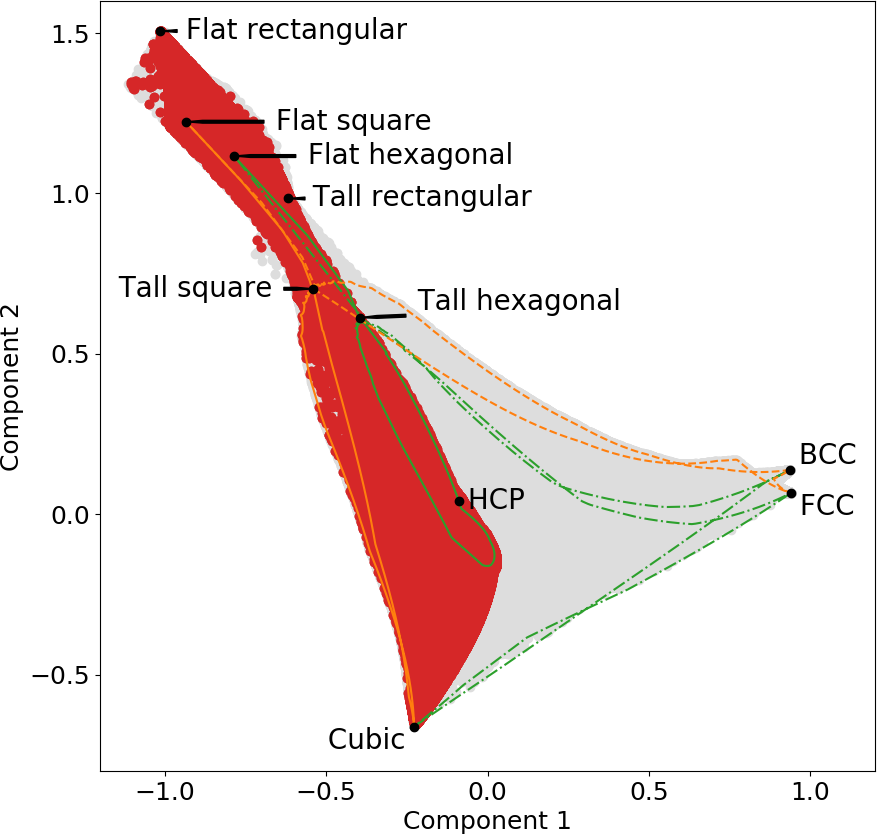}
&\includegraphics[width=0.6\columnwidth]{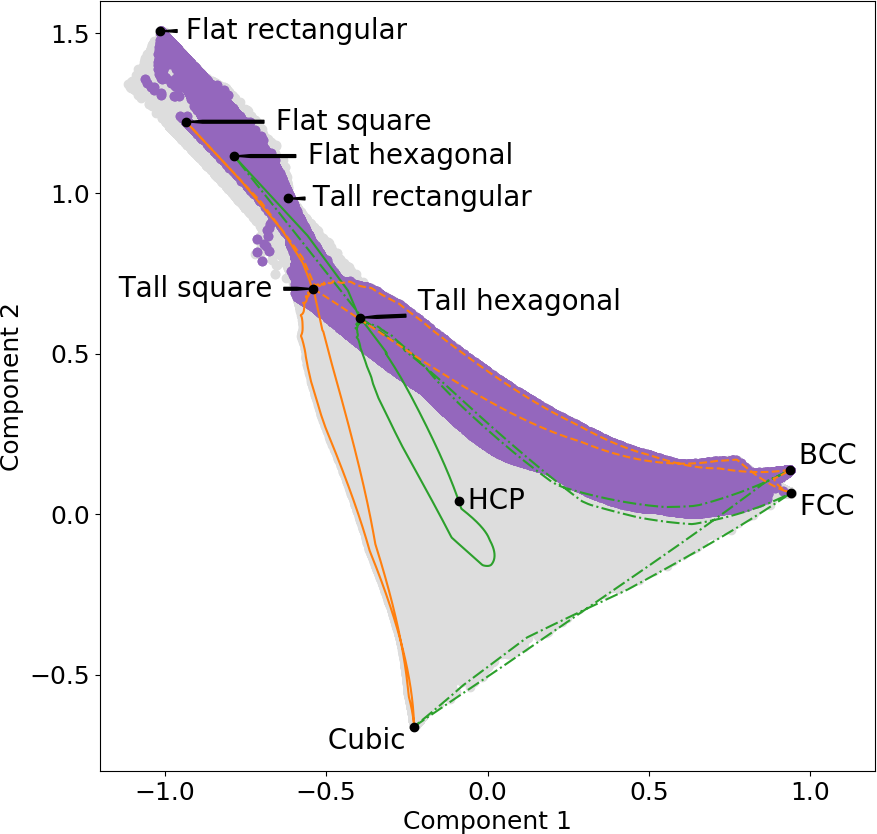}
&\includegraphics[width=0.6\columnwidth]{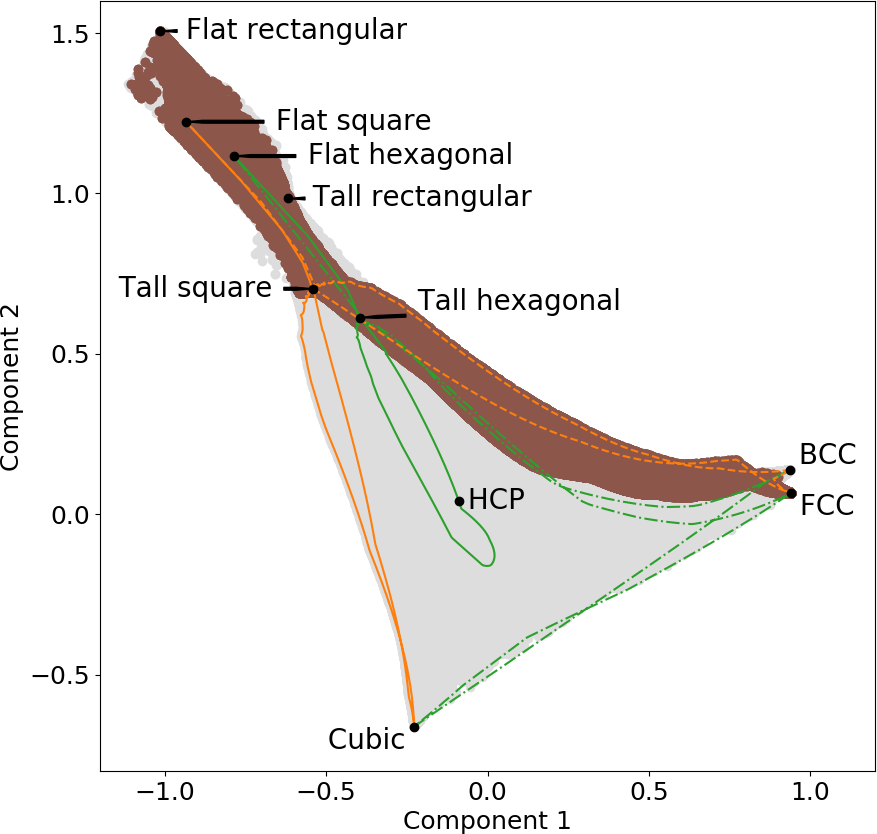}\\
\figtext{(f) Base-centred orthorhombic}
& \figtext{(g) Body-centred orthorhombic}
& \figtext{(h) Face-centred orthorhombic}
\end{tabular}
\vspace{5mm}\\
\figtext{Two-parameter Bravais lattices:}\\
\begin{tabular}{ll}
\includegraphics[width=0.08\textwidth]{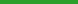} \figtext{Primitive hexagonal}\hspace{3mm}
&\includegraphics[width=0.08\textwidth]{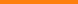} \figtext{Primitive tetragonal}\hspace{3mm}
\\
\includegraphics[width=0.08\textwidth]{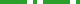} \figtext{Primitive rhombohedral}\hspace{3mm}
&\includegraphics[width=0.08\textwidth]{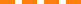} \figtext{Body-centred tetragonal}\hspace{3mm}
\end{tabular}

\caption{Two-dimensional maps of the space of Bravais lattices, obtained using PCA projections of the 14-dimensional distance vector space.  Theoretical, randomly sampled structures are shown in grey.    Experimentally observed lattices from the Crystallography Open Database (COD) are marked with blue points.   \textbf{(a)} Projection matrix is calculated using randomly sampled structures, including many degenerate Bravais lattices.  \textbf{(b)} Projection matrix is calculated using COD structures and excludes degenerate structures.  \textbf{(c)-(h)} Regions occupied by each of the monoclinic and orthorhombic lattice types in the COD projection.
\label{fig:pca_projections}}
\end{figure*}

\section{Symmetrization Distance Vectors}
\label{sec:applications}

The procedure we have described above finds a minimum-strain symmetrization to a target Bravais type, and an associated symmetrization distance.  By calculating the symmetrization distance from every Bravais type we obtain a vector of length 14, which we will call a `distance vector'.  We can develop some intuition for the distance vectors by observing how they change in the two-parameter Bravais lattices.
Figure~\ref{fig:variation_1D} shows the distance vectors for the tetragonal Bravais lattices, along a path containing all possible permitted geometries permitted by their respective symmetry conditions.

The geometry of the primitive tetragonal lattice is the simpler of the two.  At $\phi=0$, the lattice is degenerate, with lattice parameters $c=1$ and $a=0$; we denote this lattice the `tall square', since, in the limit $\phi \rightarrow 0$, the base is square.  As mentioned above, symmetrization cannot be performed for degenerate lattices, so this degenerate state is achieved only asymptotically.
At $\phi=\pi/4$, $c=a$ and the lattice is identical to the primitive cubic lattice, which is reflected by the associated zero symmetrization distance.  At $\phi=\pi/2$, the lattice is again degenerate, with lattice parameters $c=0$ and $a=1$; we denote this lattice the `flat square'.  In the interval $[\pi/2, \pi]$, the lattice is simply the mirror image of that in the interval $[0, \pi/2]$.
The path of the body-centred tetragonal lattice is similar to that of the primitive tetragonal lattice, but rather than passing through the primitive cubic lattice, it contains the \emph{Bain} transformation~\cite{bain1924nature,nishiyama2012martensitic}, from the FCC to the BCC lattice

The distances from the primitive triclinic lattice type are zero everywhere, since all lattices are trivially triclinic.  Similarly, it can be seen that the symmetrization distance for rhombohedral symmetrization is never greater than the FCC distance.  Both lattice types have lattice vectors of equal length, but the rhombohedral lattice has a variable angle.  The extra degree of freedom serves to reduce the strain.  This illustrates the important concept that Bravais lattices exist in a hierarchy, whereby the symmetries of some Bravais types are subsets of other types.
The subset relationships are illustrated in Figure~\ref{fig:hierarchy}.  In general we can state: for two lattice types, $B$ and $B^\prime$, if the symmetries of $B$ are a subset of those of $B^\prime$, then
\begin{equation}
\min_{\set{Z} \in \vectorspace{R}{3 \times 3}} d_B \left( \set{Z}, \set{B} \right)
\leq \min_{\set{Z} \in \vectorspace{R}{3 \times 3}} d_{B^\prime} \left( \set{Z}, \set{B} \right)
\end{equation}
This fact should be considered when classifying a lattice.  After imposing a threshold on the symmetrization distance, multiple potential Bravais types can be assigned.  In this case, the Bravais type with the highest number of symmetries should be selected.

\section{A Map of the Bravais Lattices}
\label{sec:cartography}

As shown above, by calculating the symmetrization distance from each Bravais type, we can assign every lattice a 14-dimensional distance vector.  It is instructive to consider the space of possible distance vectors.  To do so we use truncated principal component analysis (PCA).  Figure~\ref{fig:pca_projections} shows a two-dimensional PCA projection of distance vectors.  Two projections have been used to illustrate the distance vector space.  The PCA projection in Figure~\ref{fig:pca_projections}a is calculated using the randomly sampled Bravais lattices, including many degenerate lattices.  Whilst this projection is useful for studying the complete space of Bravais geometries, it wastes a lot of space on physically uninteresting lattices with degenerate geometries.  The PCA projection in Figures~\ref{fig:pca_projections}b-\ref{fig:pca_projections}h is calculated using structures from the Crystallography Open Database~\cite{grazulis2012cod} (COD).  Since only experimentally observed structures are used (as filtered in earlier work by some of the authors~\cite{larsen2019lowdim}), the region of interest for practical applications occupies a significantly larger area.

The extreme vertices in Figure~\ref{fig:pca_projections}a consist of the three cubic Bravais types and six degenerate Bravais lattices.  Within each of the cubic Bravais types, lattices differ by a scale factor only.  Since the symmetrization distance is invariant to scale, all lattices within a cubic Bravais type have the same distance vector.  Four of the other six extreme vertices are degenerate states of the primitive tetragonal and primitive hexagonal Bravais lattices: the `tall square', and `flat square' lattices (described above), and the `tall hexagonal' and `flat hexagonal' lattices, which are similar to the square variants but with hexagonal bases.  The two remaining extreme vertices, the `flat rectangular' and `tall rectangular' lattices, are degenerate states of the primitive orthorhombic lattice, whose cell vectors have sufficiently different lengths that two degrees of freedom are insufficient to accommodate a low-strain symmetrization.

The tetragonal and hexagonal Bravais types have two degrees of freedom.  As shown above, the scale invariance of the symmetrization distance means that the distance vectors of these Bravais types exist on a line.  The Bain transformation (from FCC to BCC), which can be achieved by tracing the body-centred tetragonal transition line, is widely studied due to its energetic accessibility in iron-based materials.  It is noteworthy that the small strain required for this transformation is reflected by the proximity of the FCC and BCC distance vectors.

\begin{figure}
\centering
\includegraphics[width=0.95\columnwidth]{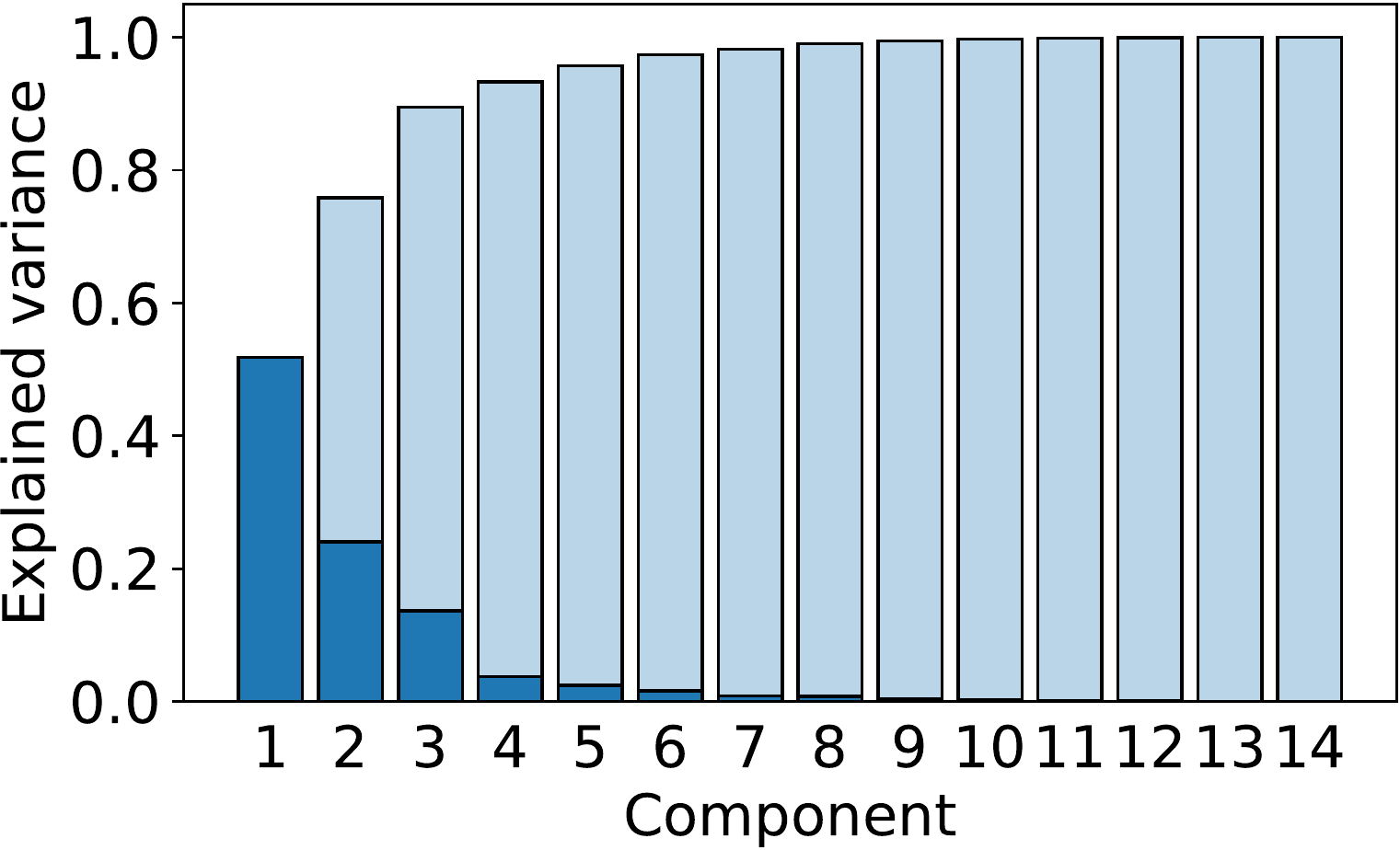}
\caption{Fraction of variance explained by each PCA component (dark blue), and the cumulative sum of the explained variance (light blue).  The projection matrix is calculated using experimental structures taken from the Crystallography Open Database.
\label{fig:pca_variance}}
\end{figure}

The orthorhombic lattices have three degrees of freedom.  Distance vectors of these lattice types exist on a two-dimensional manifold.  The monoclinic lattices have four degrees of freedom and exist on a three-dimensional manifold.
The two-dimensional PCA projection captures the symmetry subset relationships shown in Figure~\ref{fig:hierarchy}, but also introduces a degree of spurious overlap.  The overlap is worse for lattices with more degrees of freedom.  For example, the path traced by body-centred tetragonal lattice appears to self-intersect at multiple points.  These crossings are resolved in a three-dimensional projection.  Similarly, the spaces occupied by the primitive and base-centred monoclinic lattices appear to overlap, despite being disjoint.

To explain the overlap, we observe that five parameters are sufficient to describe a Bravais lattice if scale is ignored, and that the space of distance vectors is therefore a non-linear five-dimensional manifold embedded in $\vectorspace{R}{14}$.  It is therefore inevitable that a two-dimensional projection contains spurious overlap.  The loss of information resulting from PCA projection is shown in Figure~\ref{fig:pca_variance}.  PCA is a linear projection and therefore requires more than five dimensions to capture the full distance vector space.  Nonetheless, even a two-dimensional projection explains most of the variance and is useful for visualizing the space of lattices, and particularly for studying phase transitions.

Although the HCP crystal structure is not a Bravais lattice, we have marked it on the projection for reference.  In structure maps based on energetic descriptors~\cite{jenke2018descriptor}, the HCP and FCC structures are typically close together, due to their small energetic differences.  Geometrically, however, the primitive hexagonal and FCC lattices are far apart, since a large strain is required to transform one lattice into the other.  The difference between an energetic map and our map can be described in terms of transformation paths: an energetic map encodes differences in the energetic \emph{endpoints} of the transition, whereas our map encodes the lattice deformation of the transition itself.

\section{Conclusion}
\label{sec:conclusion}

We have described a method for symmetrization of Bravais lattices using strain, which is a physically intuitive quantity that also has the attractive features of rotation and scale invariance.  A distance from each Bravais type is determined by quantifying the strain necessary for symmetrization.  This allows classification to be performed without the need for an \emph{a priori} selection of a tolerance parameter.

By projecting the distance vectors using PCA, we obtain insight into the positions of lattices in an abstract Bravais space, where distances are determined by displacive deformations only.  Similarly, the symmetrization procedure finds minimum-strain solutions, rather than minimum-energy solutions.  Indeed, the symmetrization effectively operates in a continuum model which does not consider atoms at all.  Nonetheless, we envisage that the method will be useful for a range of experimental and computational applications.  

A software implementation (C\texttt{++} with Python wrappers) is available online~\cite{augustegithub}.

\begin{acknowledgments}
This work was supported by Grant No. 7026-00126B from the Danish Council for Independent Research.
\end{acknowledgments}

\appendix*
\section{Numerical Solution}
Here we rework the distance expression in Equation~\vareqref{var:objective} into a multivariate polynomial.  We perform a substitution
\begin{equation}
\set{Z} = \set{Q} \sum\limits_{i=1}^n x_i \set{T}_i
\end{equation}
where $\set{Q} \in SO(3)$ is a right-handed orthogonal matrix, and $\left\{ \set{T}_i \mid i \in 1\ldots n \right\}$ is a set of $n$ matrices (a `template') which, in conjunction with a vector $\vec{x} \in \vectorspace{R}{n}$, parametrizes a Bravais lattice type.  The templates maintains the desired symmetry by construction.  For example, a primitive orthorhombic basis has a template
\begin{equation}
\set{T}_1 = 
\begin{bmatrix}
1 & 0 & 0\\
0 & 0 & 0\\
0 & 0 & 0\\
\end{bmatrix}
\set{T}_2 = 
\begin{bmatrix}
0 & 0 & 0\\
0 & 1 & 0\\
0 & 0 & 0\\
\end{bmatrix}
\set{T}_3 = 
\begin{bmatrix}
0 & 0 & 0\\
0 & 0 & 0\\
0 & 0 & 1\\
\end{bmatrix}
\end{equation}
A primitive rhombohedral lattice has a template
\begin{equation}
\set{T}_1 = 
\begin{bmatrix}
1 & 0 & 0\\
0 & 1 & 0\\
0 & 0 & 1\\
\end{bmatrix}
\set{T}_2 = 
\begin{bmatrix}
0 & 1 & 1\\
1 & 0 & 1\\
1 & 1 & 0\\
\end{bmatrix}
\end{equation}
Similar templates can be constructed for the other Bravais lattice types.

\begin{figure}[!h]
\centering
\includegraphics[width=0.45\columnwidth]{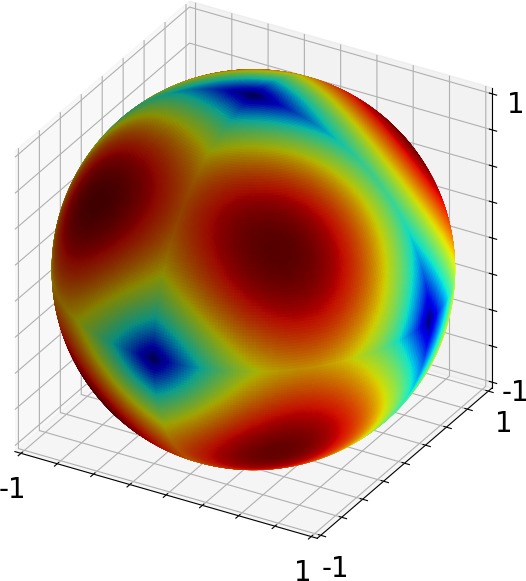}
\caption{Function landscape for a primitive orthorhombic cell, after application of the Mahalonobis transformation.  The axes are the cell parameters $\vec{x}$.  At each point the optimal orthogonal component (\set{Q}) has been determined.  Due to application of the Mahalonobis transformation the maxima (shown in red) are symmetric and have identical function values.
\label{fig:global_optima}}
\end{figure}

In order to express the distance function as a multivariate polynomial we parametrize the orthogonal matrix using quaternions.
Briefly, quaternions are numbers of the form $\vec{q} = [q_1, \textbf{i} q_2, \textbf{j} q_3, \textbf{k} q_4]$ which generalize the complex numbers~\cite{altmann2005rotations}.  Unit quaternions are a useful parametrization of $SO(3)$, the space of rotations.  The quaternion-derived rotation matrix is given by:
\begin{equation}
\set{Q} = 
\begin{bmatrix}
1 {-2q_3^2} {-2q_4^2}
&2q_2q_3 - 2q_1q_4
&2q_2q_4 + 2q_1q_3\\
2q_2q_3 + 2q_1q_4
&1 {-2q_2^2} {-2q_4^2}
&2q_3q_4 - 2q_1q_2\\
2q_2q_4 - 2q_1q_3
&2q_3q_4 + 2q_1q_2
&1 {-2q_2^2} {-2q_3^2}
\end{bmatrix}
\end{equation}
The distance function can then be expressed as
\begin{equation}
3 + \frobnorm{ \sum\limits_{i=1}^n x_i \set{T}_i \set{B}^{-1} }^2 - 2\Tr \left( \set{Q} \sum\limits_{i=1}^n x_i \set{T}_i \set{B}^{-1} \right)
\end{equation}
which is a multivariate polynomial of degree six.  We can reduce the degree with a variable substitution.
Let $\set{H} \in \vectorspace{R}{n \times n}$ be the symmetric matrix with elements
\begin{equation}
\set{H}_{ij} = \frobproduct{ \set{T}_i \set{B}^{-1}, \set{T}_j \set{B}^{-1} }
\end{equation}
Then \set{H} is positive semidefinite by construction and the eigenvalues of \set{H} are non-negative.
As such, we can perform a Mahalonobis decomposition
\begin{equation}
\set{H}^{-1/2} = \set{\Gamma} \set{\Lambda}^{-1/2} \set{\Gamma}^T
\end{equation}
also known as ZCA whitening or sphering~\cite{bell1997zca,li1998sphering}.  
Here $\set{\Gamma}$ is an orthogonal matrix whose columns are the eigenvectors of \set{H} and $\set{\Lambda} = \diag \left( \lambda_1, \lambda_2, \ldots, \lambda_n \right) $ is a diagonal matrix of the eigenvalues of \set{H}.
Per the definition of \set{H}
\begin{equation}
\frobnorm{ \sum\limits_{i=1}^n x_i \set{T}_i \set{B}^{-1} }^2 = \vec{x}^T \set{H} \vec{x}
\end{equation}
and, therefore
\begin{equation}
\frobnorm{ \sum\limits_{i=1}^n \left( \set{H}^{-1/2} \vec{x} \right)_i \set{T}_i \set{B}^{-1} } = 1
\hspace{3mm} \forall \vec{x} : \norm{ \vec{x} } = 1
\end{equation}
With this transformation made we can solve an equivalent constrained problem of degree three:
\begin{equation}
\begin{split}
\max_{ \set{q} \in \vectorspace{R}{4}, \vec{x} \in \vectorspace{R}{n} }
\Tr \left( \set{Q} \sum\limits_{i=1}^n \left( \set{H}^{-1/2} \vec{x} \right)_i \set{T}_i \set{B}^{-1} \right)\\
\text{subject to:} \hspace{4mm} \norm{ \vec{q} } = 1
\hspace{4mm}
\norm{ \vec{x} } = 1
\end{split}
\label{eq:constrained}
\end{equation}
An appropriate scaling of $\vec{x}$ can be found post-solution.
A consequence of the Mahalonobis transformation is that all maxima of this equation are symmetrically equivalent and have equal solution values (c.f.~Figure~\ref{fig:global_optima}).

Equation~(\ref{eq:constrained}) can be solved using stepwise iteration~\cite{orthogonal1998everson}.  This is guaranteed to converge to a local maximum (rather than a saddle point) but has only at a linear convergence rate.  On the other hand, by solving the Lagrangian form of Equation~(\ref{eq:constrained}) we can use Newton-Raphson iteration to achieve quadratic convergence.  We have found that a good compromise is to perform stepwise iteration for 10-20 iterations (to get close to the maximum), and then switch to Newton-Raphson (to get fast convergence).

The total number of iterations required is dependent on the number of template variables, but even for monoclinic templates convergence is typically achieved within 20 iterations.  For cubic templates the problem reduces to a polar decomposition, and convergence is achieved in a single stepwise iteration.

%

\end{document}